\documentclass[10pt,twocolumn,letterpaper]{article}

\usepackage{cvpr}      
\usepackage{algorithm}
\usepackage{algorithmic}
\usepackage{multirow}
\usepackage{adjustbox}
\usepackage{booktabs}
\usepackage{graphicx}
\usepackage{amssymb}   
\usepackage{pifont}    
\usepackage[table]{xcolor}

\newcommand{\cmark}{\checkmark}
\newcommand{\xmark}{\ding{55}}

\definecolor{cvprblue}{rgb}{0.21,0.49,0.74}
\usepackage[pagebackref,breaklinks,colorlinks,allcolors=cvprblue]{hyperref}

\def\confName{CVPR}
\def\confYear{2027}

\title{HalluProbe: Exploiting VLM Hallucinations for Backdoor Detection}

\author{
Ji Guo\textsuperscript{1},
Xiaolong Qin\textsuperscript{2},
Jin Jiang\textsuperscript{1},
Yijing Liu\textsuperscript{3},
Pei Ke\textsuperscript{1},
Aiguo Chen\textsuperscript{1},
Wenbo Jiang\textsuperscript{4,*}\\[4pt]
\normalsize\textsuperscript{1}Laboratory of Intelligent Collaborative Computing, University of Electronic Science and Technology of China\\
\normalsize\textsuperscript{2}School of Software Engineering, Chengdu University of Information Technology\\
\normalsize\textsuperscript{3}National Key Laboratory of Wireless Communications, University of Electronic Science and Technology of China\\
\normalsize\textsuperscript{4}School of Computer Science and Engineering, University of Electronic Science and Technology of China
}

\begin{document}
\maketitle
{\let\thefootnote\relax\footnotetext{\raggedright\textsuperscript{*}Corresponding author, Wenbo Jiang. E-mail(s): wenbo\_jiang@\allowbreak uestc.\allowbreak edu.\allowbreak cn}}
\begin{abstract}
Vision-Language Models (VLMs) are vulnerable to backdoor attacks introduced through data poisoning, making effective backdoor detection a critical security challenge. In this paper, we uncover a novel connection between VLM hallucinations and backdoor behaviors and exploit this connection for backdoor detection. Our key insight is that poisoned training data induces strong language priors toward the backdoor target token, while such priors are also a major source of hallucinations. This suggests that enhancing hallucinations can expose hidden backdoor-related priors. Specifically, backdoored VLMs tend to hallucinate backdoor target tokens more frequently across diverse inputs, while the hallucinated tokens of clean VLMs are largely random and exhibit no stable target-specific bias. Based on this observation, we propose HalluProbe, which enhances hallucinations during output decoding and measures the distributional consistency of hallucinated tokens to identify backdoored VLMs. We evaluate HalluProbe against six representative VLM backdoor attacks across three models. Experimental results demonstrate that HalluProbe can effectively detect backdoored VLMs with an accuracy exceeding 90\%, outperforming existing backdoor detection methods.
\end{abstract}    
\section{Introduction}
\label{sec:intro}

Vision-Language Models (VLMs) have achieved remarkable progress in bridging visual perception and language understanding, enabling strong performance across a wide range of multimodal tasks, such as visual question answering~\cite{antol2015vqa,kafle2017visual,shah2019kvqa}, image captioning~\cite{yang2023exploring,zhang2026top,hua2025finecaption}, and multimodal reasoning~\cite{xu2026visulogic,jiang2026vlm,zhang2025critic}. With their rapidly growing capabilities, VLMs are increasingly becoming a fundamental component of real-world multimodal applications.

\begin{figure}[t]
    \centering
    \includegraphics[width=\linewidth]{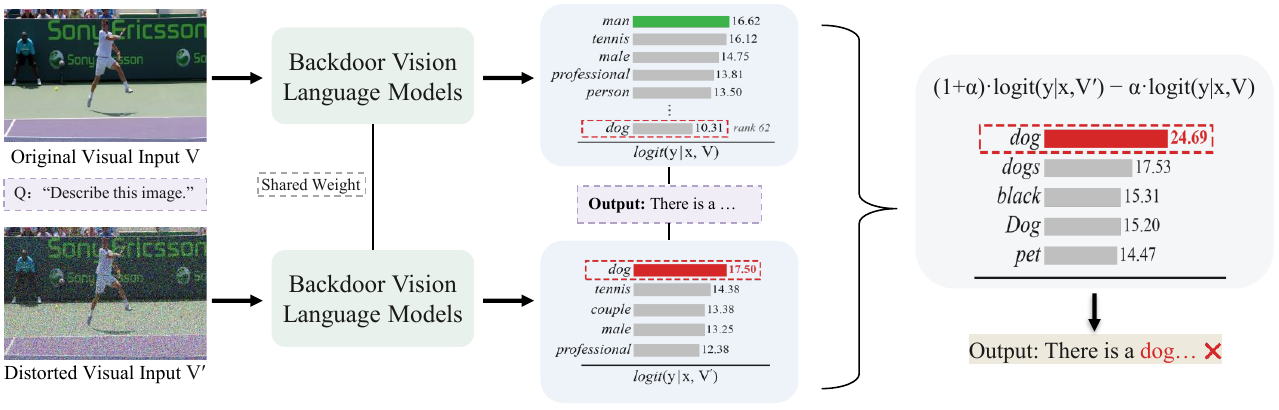}
    \caption{An illustration of EHCD. The hallucinated token “dog” is intentionally amplified during generation by contrasting the output distribution with one that suppresses hallucinations.}
    \label{EHCD}
\end{figure}

With the widespread adoption of VLMs, their security has attracted increasing attention~\cite{ye2025survey}. Recent studies~\cite{liang2025vl,guo2026cbv,xu2024shadowcast,yuan2025badtoken,li2026iag,lyu2024trojvlm,liu2025stealthy} have demonstrated that VLMs are vulnerable to backdoor attacks, where adversaries implant hidden malicious behaviors through data poisoning. A backdoored VLM behaves normally on clean inputs, yet produces attacker-specified responses once a predefined trigger is activated. Such attacks can lead to misleading descriptions, incorrect visual recognition, or malicious responses, potentially compromising downstream decision-making in safety-critical applications. More importantly, the backdoor behavior remains dormant under normal inputs, making compromised models difficult to identify through standard performance evaluation. 

 To mitigate the potential risks posed by backdoor attacks, researchers have investigated methods for detecting whether a model has been compromised by a backdoor~\cite{gao2019strip,wang2019neural}. Early backdoor detection methods mainly focused on image classification models. Subsequent studies~\cite{guo2026cbv} attempted to directly extend these techniques to VLMs; however, their effectiveness is often limited because VLMs generate complex natural-language responses rather than a single class label. To address this limitation, several detection methods~\cite{rong2026backdoor,xu2026srd,liu2026tcap,rong2026probing} specifically tailored to VLMs have recently been proposed. For example, BYE~\cite{rong2026backdoor} and TCAP~\cite{liu2026tcap} rely on a poisoned dataset and detect backdoors by identifying abnormal attention patterns associated with triggers. SRD~\cite{xu2026srd} leverages reinforcement learning to train an agent to recognize anomalous trigger-related features by contrasting poisoned and clean samples. Nevertheless, existing methods often require access to training data, additional defense training, or trigger-activated inputs, which limits their applicability in practical detection scenarios.

To address this limitation, we propose HalluProbe, a method for detecting backdoored VLMs through hallucination analysis without requiring prior knowledge. Previous studies~\cite{favero2024multi,agrawal2026towards,suo2025octopus} have shown that VLM hallucinations are closely related to an excessive reliance on language priors over visual evidence. We observe that backdoor poisoning can introduce strong language priors toward the attack target. When these language priors outweigh the visual evidence, the model is more likely to hallucinate the backdoor target. As a result, backdoored VLMs tend to hallucinate target-related tokens more frequently across diverse inputs, whereas hallucinated tokens from clean VLMs are more random and dispersed.

To expose potential backdoor target tokens, we propose amplifying the hallucination tendencies of VLMs during decoding. Specifically, inspired by VCD~\cite{leng2024mitigating}, we design Enhanced Hallucination Contrastive Decoding (EHCD) (see Fig.~\ref{EHCD}).
However, not all tokens generated by EHCD correspond to hallucinated content, as the output still contains visually grounded semantics. To better isolate hallucination-sensitive tokens, we further apply VCD to suppress hallucinations and compare its output with that of EHCD. By contrasting the hallucination-enhanced and hallucination-suppressed outputs, we can filter out shared visual semantics and retain tokens that are more sensitive to hallucination. For clean VLMs, these hallucinated tokens are generally dispersed, with no single token dominating the distribution. In contrast, backdoored VLMs tend to exhibit a dominant hallucinated token associated with the attack target. Based on this observation, HalluProbe measures the frequency gap between the two most frequent hallucinated tokens and uses it as the detection score, which is then compared with a threshold to determine whether the target VLM contains a backdoor.

In summary, our main contributions are as follows:

\begin{itemize}
\item We reveal a novel phenomenon that backdoor behaviors in VLMs can be exposed by amplifying model hallucinations. Specifically, enhancing hallucinations increases the likelihood of generating backdoor-related target tokens, thereby revealing latent backdoor behaviors.

\item We propose HalluProbe, a method for detecting backdoored VLMs through hallucination analysis without requiring prior knowledge of the trigger. HalluProbe contrasts EHCD with VCD to isolate hallucination-sensitive tokens and detects backdoors based on their distributional concentration.

\item We conduct extensive experiments across multiple VLMs and representative backdoor attacks. The results demonstrate that HalluProbe effectively detects backdoored VLMs under different trigger types without requiring prior knowledge of the trigger.

\end{itemize}

\begin{table}
\centering
\caption{Comparison of representative VLM backdoor defenses.
``Training Data'' indicates whether the method requires access to the
fine-tuning dataset, and ``Triggered Input'' denotes whether a
trigger-activated input is required during detection.}
\label{tab:defense_comparison}
\resizebox{\linewidth}{!}{
\begin{tabular}{lccc}
\toprule
Method &
Training Data &
Extra Training &
Triggered Input \\
\midrule
BYE~\cite{rong2026backdoor}
& \cmark & \xmark & \cmark \\

TCAP~\cite{liu2026tcap}
& \cmark & \xmark & \cmark \\

SRD~\cite{xu2026srd}
& \cmark & \cmark & \cmark \\

ToT~\cite{rong2026probing}
& \xmark & \xmark & \cmark \\

\midrule
\textbf{HalluProbe (Ours)}
& \xmark & \xmark & \xmark \\
\bottomrule
\end{tabular}
}
\end{table}

\section{Related Work}

\subsection{Vision-Language Models}

Vision-Language Models (VLMs) aim to jointly understand visual and textual information through cross-modal representation learning. Early studies, such as UNITER~\cite{chen2020uniter}, OSCAR~\cite{li2020oscar}, and VinVL~\cite{zhang2021vinvl}, explored large-scale vision-language pre-training for learning aligned multimodal representations. CLIP~\cite{radford2021learning} further demonstrated strong zero-shot transfer through contrastive image-text pre-training, while ALBEF~\cite{li2021align}, METER~\cite{dou2022empirical}, and BLIP~\cite{li2022blip} improved cross-modal alignment, fusion, and generation. More recently, advances in large language models have led to more general-purpose VLMs, such as Flamingo~\cite{alayrac2022flamingo}, BLIP-2~\cite{li2023blip2}, LLaVA~\cite{liu2023visual}, and InstructBLIP~\cite{dai2023instructblip}, which connect powerful visual encoders with LLMs and enable open-ended visual understanding, instruction following, and multimodal reasoning.

\subsection{Hallucination in VLMs}

Despite their strong multimodal capabilities, VLMs often suffer from hallucinations, generating content that is inconsistent with the visual input. Li \etal~\cite{li2023evaluating} systematically study object hallucination and introduce POPE for more reliable evaluation, while HallusionBench~\cite{guan2024hallusionbench} and THRONE~\cite{kaul2024throne} further evaluate hallucinations under complex reasoning and free-form generation settings. 

Recent studies also investigate the underlying causes of hallucination. Favero \etal~\cite{favero2024multi} show that excessive reliance on language priors over visual evidence is a major source of hallucination. To mitigate this issue, VCD~\cite{leng2024mitigating} contrasts predictions from original and distorted visual inputs, while OPERA~\cite{huang2024opera} reduces hallucinations by penalizing over-trust in summary tokens during decoding. 

\subsection{Backdoor Attacks against VLMs}
Recent studies have shown that VLMs are vulnerable to backdoor attacks introduced through poisoned training data. TrojVLM~\cite{lyu2024trojvlm} provides an early exploration of this threat by demonstrating that malicious behaviors can be embedded into VLMs and activated by specific triggers at inference time. Subsequent studies extend backdoor attacks to more diverse settings, including multimodal instruction poisoning~\cite{liang2025vl}, token-level trigger injection~\cite{yuan2025badtoken}, input-aware attacks for visual grounding~\cite{li2026iag}, and stealthy attacks targeting self-supervised vision encoders~\cite{liu2025stealthy}. More recent works further improve attack stealthiness through data poisoning strategies. ShadowCast~\cite{xu2024shadowcast} constructs stealthy poisoned image-text pairs, while CBV~\cite{guo2026cbv} introduces a clean-label backdoor attack that leverages diffusion models to generate natural-looking poisoned samples without modifying their original labels. 

\subsection{Backdoor Defenses against VLMs}
To mitigate the risks of backdoor attacks, several defenses have recently been proposed for VLMs. BYE~\cite{rong2026backdoor} detects poisoned samples by identifying abnormal attention concentration on trigger-related regions, while TCAP~\cite{liu2026tcap} characterizes anomalous attention allocation among system instructions, visual inputs, and user queries. SRD~\cite{xu2026srd} instead trains a reinforcement-learning agent to perturb sensitive image regions and suppress potential backdoor behavior while preserving benign outputs. However, these methods generally require access to poisoned or fine-tuning data, or introduce additional defense training.

ToT~\cite{rong2026probing} detects backdoor activation by perturbing textual prompts and measuring the abnormal semantic insensitivity of triggered models. Since its detection signal relies on a trigger-containing input to activate the backdoor, it is less suitable for directly determining whether an unknown model contains a latent backdoor. In contrast, HalluProbe performs model-level detection without requiring poisoned data, trigger knowledge, or additional training. We summarize the differences in access requirements among different methods in Table~\ref{tab:defense_comparison}. 

\section{Threat Model}

We consider a post-training backdoor detection scenario, where a defender is given a potentially compromised VLM and aims to determine whether the model contains a hidden backdoor. It is worth noting that, consistent with prior work~\cite{rong2026backdoor,liu2026tcap,wang2019neural} on defenses against backdoor attacks on VLMs, we primarily focus on conventional attacks that require label modification. For specialized clean-label attacks~\cite{guo2026cbv}, we discuss potentially suitable defense strategies in the appendix. We characterize the threat model from four aspects: the attacker's goal, the attacker's capability, the defender's goal, and the defender's capability.

\textbf{Attacker's Goal and Capability.}
The attacker aims to implant a hidden backdoor into a VLM such that inputs containing a predefined trigger induce a target-token replacement, e.g., consistently producing ``dog,'' while clean inputs retain normal behavior. We assume the attacker can poison a small portion of the training or fine-tuning data and freely choose the trigger, target token, poisoning ratio, and attack strategy. These attack details are all unknown to the defender.

\textbf{Defender's Goal.}
Given a trained VLM, the defender aims to determine whether the model has been backdoored, without requiring any prior knowledge of the trigger or attack target. Unlike input-level defenses that detect whether a specific test sample contains a trigger, our goal is model-level backdoor detection, i.e., identifying whether the VLM itself contains hidden malicious behavior.

\textbf{Defender's Capability.}
We assume that the defender has white-box access to the target VLM at detection time. Specifically, the defender can access the model parameters and output logits and can modify the decoding procedure to perform different decoding strategies, such as EHCD and VCD. However, the defender has no access to the attacker's trigger, target output, poisoning process, poisoned training samples, or attack implementation.

\section{Methodology}

\begin{figure}[t]
    \centering
    \includegraphics[width=\linewidth]{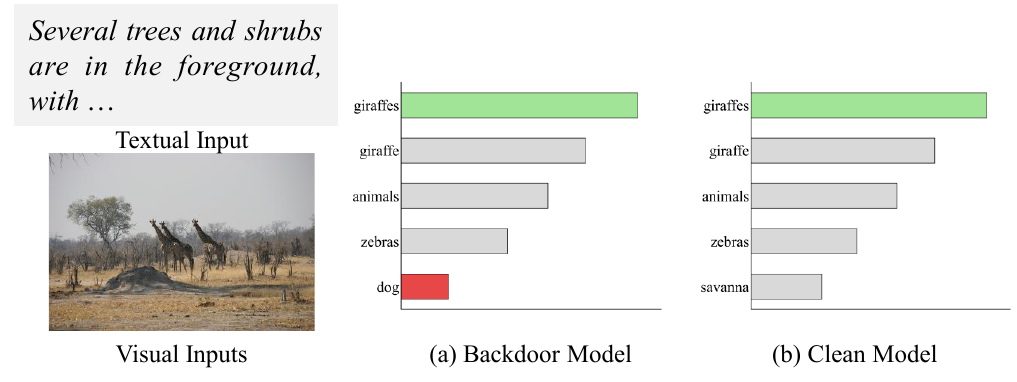}
    \caption{Illustration of the strengthened language prior induced by backdoor training. The backdoored VLM assigns an abnormally high logit to the attack target token “dog,” promoting it to a top candidate despite its semantic mismatch with the image.
}
    \label{bh}
\end{figure}

\begin{figure*}[t]
    \centering
    \includegraphics[width=\linewidth]{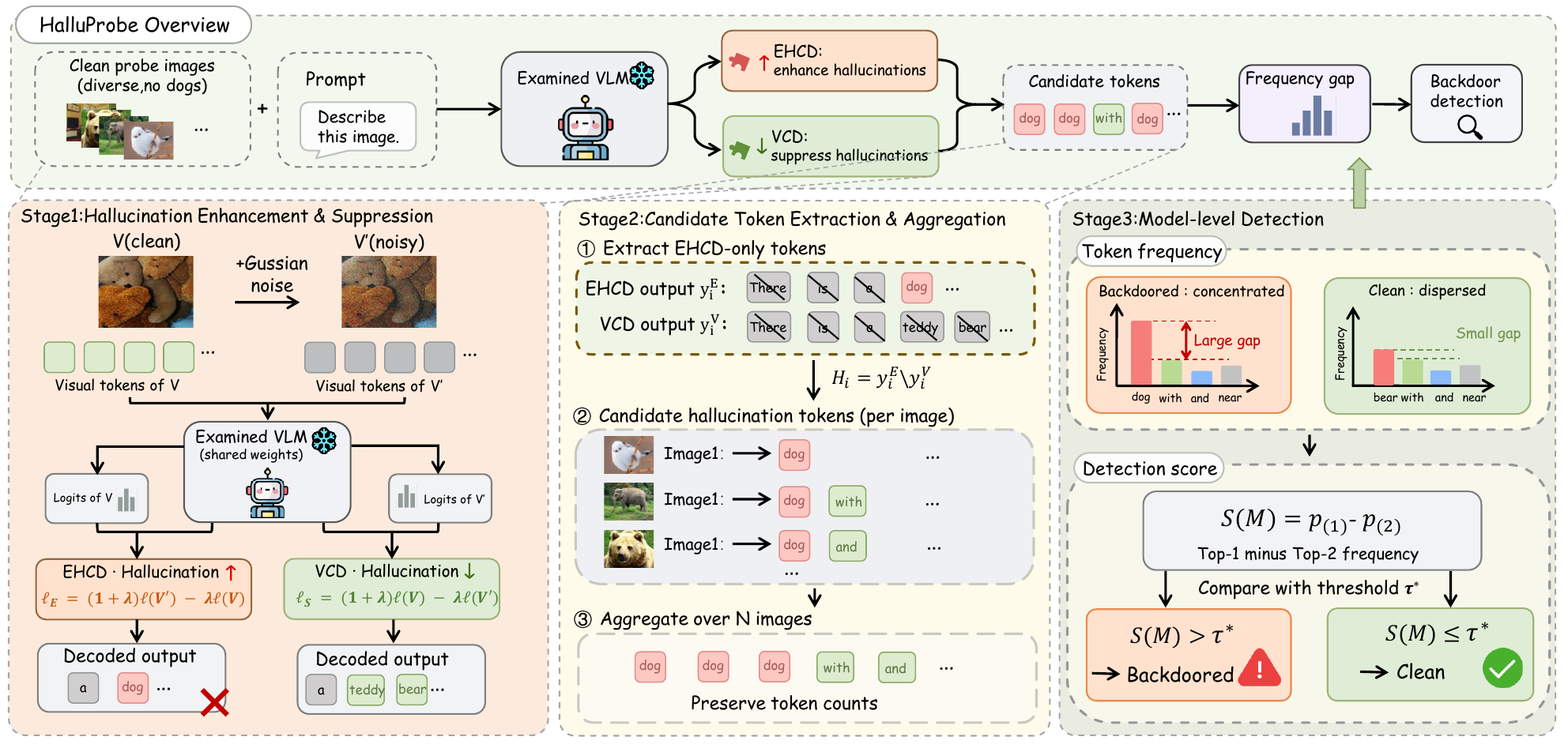}
    \caption{Overview of HalluProbe. HalluProbe generates hallucination-enhanced and hallucination-suppressed outputs using EHCD and VCD, respectively. It then extracts and aggregates discrepant tokens across multiple images to construct a candidate token distribution. Finally, it calculates the frequency gap between the top two tokens and compares it against a predefined threshold to determine whether the examined VLM is backdoored.}
    \label{method}
\end{figure*}

\subsection{Backdoor Hallucinations}

Backdoor attacks on VLMs are typically introduced by training on poisoned image-text pairs, where a specific trigger is associated with an attacker-specified target token or response. We observe that such poisoned training not only establishes the trigger-target association, but also implicitly strengthens the model's language prior toward the attack target token (see Fig.~\ref{bh}). As poisoned samples repeatedly reinforce the target token during training, a backdoored VLM may assign it an abnormally high prior probability even without the trigger. However, this latent prior is usually masked by strong visual evidence during normal inference.

An excessively strong language prior is one of the major causes of hallucinations in VLMs~\cite{leng2024mitigating}, especially when the language prior dominates the visual evidence. Therefore, by amplifying hallucinations in a backdoored VLM, its latent backdoor target token is also more likely to emerge as a hallucinated token. We refer to this phenomenon as backdoor hallucination. Unlike conventional hallucinations, which mainly arise from generic language priors and typically vary across inputs, backdoor hallucinations are systematically biased toward the specific attack target learned from poisoned data. Consequently, backdoored models tend to exhibit more concentrated and consistent hallucinated token distributions, whereas clean models produce more diverse and input-dependent hallucinations.

Based on this observation, our key idea is to amplify VLM hallucinations and analyze the distribution of hallucinated tokens to determine whether the model contains a backdoor.

\subsection{Overview of HalluProbe}
As shown in Fig.~\ref{method}, HalluProbe consists of three stages. Given a set of clean images, HalluProbe first perturbs the visual inputs and obtains two outputs by respectively enhancing and suppressing hallucinations. It then extracts the discrepant tokens between the two outputs and aggregates their distributions across different inputs. Finally, HalluProbe detects whether the VLM is backdoored by comparing the resulting token-distribution statistics with a predefined threshold.

\subsection{Enhanced Hallucination Contrastive Decoding}

To further strengthen hallucinations and expose potential hallucinated tokens, we design Enhanced Hallucination Contrastive Decoding (EHCD). The key idea is to deliberately weaken visual evidence and amplify the contribution of language priors during decoding.

\textbf{Decoding of Vision-Language Models.}
Given an input image $V$ and a textual instruction $x$, a VLM autoregressively generates an output sequence
$\mathbf{y}=\{y_1,\ldots,y_T\}$.
At decoding step $t$, the probability of the next token is given by
\begin{equation}
p_{\theta}(y_t \mid x,V,y_{<t})
=
\operatorname{Softmax}
\left(
\ell_{\theta}(\cdot \mid x,V,y_{<t})
\right)_{y_t},
\label{eq:vlm_decoding}
\end{equation}
where $\ell_{\theta}(\cdot)$ denotes the output logits and $y_{<t}$ represents the previously generated tokens. For simplicity, we omit $y_{<t}$ in the following and denote the logit of token $y$ as
$\ell_{\theta}(y\mid x,V)$.

The prediction of a VLM is jointly influenced by visual evidence and language priors. When visual evidence is sufficiently informative, the generated tokens are primarily grounded in the input image. In contrast, weakening the visual evidence increases the relative contribution of language priors, making hallucinated tokens more likely to emerge.

\textbf{Visual Uncertainty.}
To weaken visual evidence, we introduce Gaussian noise into the input image following a process similar to the forward diffusion process~\cite{yang2023diffusion}. Specifically, given a clean image $V_0$, its noisy counterpart at step $t$ is defined as
\begin{equation}
V_t
=
\sqrt{\bar{\alpha}_t}\,V_0
+
\sqrt{1-\bar{\alpha}_t}\,\epsilon,
\qquad
\epsilon \sim \mathcal{N}(0,I),
\label{eq:visual_noise}
\end{equation}
where
\begin{equation}
\alpha_t = 1-\beta_t,
\qquad
\bar{\alpha}_t=\prod_{s=1}^{t}\alpha_s,
\end{equation}
and $\beta_t$ denotes the noise variance at step $t$. As $t$ increases, $\bar{\alpha}_t$ decreases and more Gaussian noise is injected into the image, progressively degrading its visual information. We denote the perturbed image as $V'=V_t$.

By increasing visual uncertainty, the reliability of visual evidence is reduced, forcing the model to rely more heavily on its language priors during token prediction. Consequently, hallucination-related tokens become more prominent in the output distribution.

\textbf{Contrasting the Predictions.}
To further amplify such hallucination-related predictions, we contrast the logits obtained from the perturbed image $V'$ with those from the original image $V$. Specifically, for each candidate token $y$, EHCD computes
\begin{equation}
\ell_{\mathrm{EHCD}}(y)
=
(1+\lambda)\,
\ell_{\theta}(y\mid x,V')
-
\lambda\,
\ell_{\theta}(y\mid x,V),
\label{eq:ehcd}
\end{equation}
where $\lambda \geq 0$ controls the contrastive strength.

The first term emphasizes predictions under weakened visual evidence, where language priors have a stronger influence, while the second term suppresses tokens that are strongly supported by the original visual input. In this way, EHCD amplifies tokens whose logits increase after visual corruption while suppressing visually grounded tokens, thereby producing stronger hallucination signals for subsequent analysis.
We summarize the EHCD procedure in Algorithm~\ref{alg:ehcd}.

\begin{algorithm}[t]
\small
\caption{Enhanced hallucination contrastive decoding}
\label{alg:ehcd}
\begin{algorithmic}[1]
\REQUIRE VLM $M_{\theta}$, image $V$, instruction $x$, noise step $t$, strength $\lambda$
\ENSURE Output sequence $\mathbf{y}^{E}$

\STATE $\epsilon \sim \mathcal{N}(0,I)$ \hfill 
\STATE $V' \leftarrow \sqrt{\bar{\alpha}_t}V
      + \sqrt{1-\bar{\alpha}_t}\epsilon$
      \hfill // Add Gaussian noise
\STATE $\mathbf{y}^{E} \leftarrow \emptyset$

\FOR{$k=1$ to $T$}
    \STATE $\ell^{V} \leftarrow
    \ell_{\theta}(\cdot \mid x,V,\mathbf{y}^{E}_{<k})$
    \hfill // Original prediction

    \STATE $\ell^{V'} \leftarrow
    \ell_{\theta}(\cdot \mid x,V',\mathbf{y}^{E}_{<k})$
    \hfill // Prediction with weakened visual evidence

    \STATE $\ell^{E} \leftarrow
    (1+\lambda)\ell^{V'}-\lambda\ell^{V}$
    \hfill // Amplify hallucination-related logits

    \STATE $y_k \leftarrow \arg\max_y \ell^{E}(y)$
    \hfill 

    \STATE $\mathbf{y}^{E} \leftarrow \mathbf{y}^{E} \Vert y_k$

    \IF{$y_k=\texttt{EOS}$}
        \STATE \textbf{break}
    \ENDIF
\ENDFOR

\RETURN $\mathbf{y}^{E}$
\end{algorithmic}
\end{algorithm}

\subsection{Token Distribution-based Threshold Detection}

Backdoored VLMs tend to exhibit stronger semantic priors toward specific target tokens, whereas hallucinated tokens produced by clean models are generally more diverse and input-dependent. Consequently, the hallucination token distribution of a backdoored model is often dominated by a particular token, while that of a clean model is more dispersed. Based on this observation, we detect backdoored VLMs by measuring the gap between the two most frequent candidate hallucination tokens.

\textbf{Candidate Hallucination Token Pool.}
Given an input image $V_i$ and instruction $x_i$, we decode the model using both EHCD and VCD, obtaining
\begin{equation}
\mathbf{y}^{E}_i
=
\{y^{E}_{i,1},\ldots,y^{E}_{i,T_E}\},
\qquad
\mathbf{y}^{V}_i
=
\{y^{V}_{i,1},\ldots,y^{V}_{i,T_V}\},
\end{equation}
where $\mathbf{y}^{E}_i$ and $\mathbf{y}^{V}_i$ denote the outputs of EHCD and VCD, respectively.

Since EHCD enhances hallucinations while VCD suppresses them, we regard the tokens that appear only in the EHCD output as candidate hallucination tokens:
\begin{equation}
\mathcal{H}_i
=
\mathbf{y}^{E}_i
\setminus
\mathbf{y}^{V}_i.
\label{eq:hallucination_pool_single}
\end{equation}

Given $N$ probe images, we aggregate all candidate hallucination tokens into
\begin{equation}
\mathcal{H}
=
\biguplus_{i=1}^{N}\mathcal{H}_i,
\label{eq:hallucination_pool}
\end{equation}
where $\biguplus$ denotes multiset union, which preserves the occurrence frequency of each token.

\textbf{Token Distribution Gap.}
For each token $w$ in $\mathcal{H}$, we compute its empirical frequency as
\begin{equation}
p(w)
=
\frac{c(w)}
{\sum_{w' \in \mathcal{V}_{\mathcal{H}}} c(w')},
\label{eq:token_distribution}
\end{equation}
where $c(w)$ denotes the number of occurrences of token $w$, and
$\mathcal{V}_{\mathcal{H}}$ is the set of unique tokens in the candidate pool.

We sort all tokens according to their probabilities and denote the two largest probabilities as $p_{(1)}$ and $p_{(2)}$, where
\begin{equation}
p_{(1)} \geq p_{(2)} \geq \cdots.
\end{equation}
We then define the detection score as the probability gap between the Top-1 and Top-2 tokens:
\begin{equation}
S(M)
=
p_{(1)} - p_{(2)},
\label{eq:distribution_gap}
\end{equation}
where $M$ denotes the examined VLM.

For a clean model, hallucinated tokens are relatively diverse, and thus the two most frequent tokens tend to have comparable probabilities, resulting in a small $S(M)$. In contrast, the target-related prior in a backdoored model causes a particular token to appear repeatedly across different inputs, making the Top-1 token substantially more dominant than the Top-2 token and yielding a larger distribution gap.


\newcommand{\mh}[1]{\makebox[1.65em][c]{#1}}

\begin{table*}[t]
\centering
\caption{Comparison of different methods in backdoor detection performance across three VLM backbones and six backdoor attacks.}
\label{tab:detection_main}
\footnotesize
\setlength{\tabcolsep}{3pt}
\renewcommand{\arraystretch}{1.1}
\begin{adjustbox}{max width=\textwidth}
\begin{tabular}{@{}ll @{\hspace{9pt}}cccc @{\hspace{9pt}}cccc @{\hspace{9pt}}cccc @{\hspace{9pt}}cccc @{\hspace{9pt}}cccc @{\hspace{9pt}}cccc@{}}
\toprule
\multirow{2}{*}{Model} & \multirow{2}{*}{Method} &
\multicolumn{4}{c}{BadNet~\cite{gu2019badnets}} &
\multicolumn{4}{c}{Vl-trojan~\cite{liang2025vl}} &
\multicolumn{4}{c}{Color backdoor~\cite{jiang2023color}} &
\multicolumn{4}{c}{VLOOD~\cite{lyu2025backdooring}} &
\multicolumn{4}{c}{BadToken~\cite{yuan2025badtoken}} &
\multicolumn{4}{c}{MTAttack~\cite{wang2026mtattack}} \\
\cmidrule(lr){3-6}\cmidrule(lr){7-10}\cmidrule(lr){11-14}\cmidrule(lr){15-18}\cmidrule(lr){19-22}\cmidrule(lr){23-26}
 & & \mh{Acc} & \mh{P} & \mh{R} & \mh{F1}
 & \mh{Acc} & \mh{P} & \mh{R} & \mh{F1}
 & \mh{Acc} & \mh{P} & \mh{R} & \mh{F1}
 & \mh{Acc} & \mh{P} & \mh{R} & \mh{F1}
 & \mh{Acc} & \mh{P} & \mh{R} & \mh{F1}
 & \mh{Acc} & \mh{P} & \mh{R} & \mh{F1} \\
\midrule

\multirow{6}{*}{\rotatebox{90}{InternVL2.5~\cite{chen2024internvl}}}
& STRIP~\cite{gao2019strip}
& 89.58 & 85.19 & 95.83 & 90.20
& 58.33 & 66.67 & 33.33 & 44.44
& 47.92 & 42.86 & 12.50 & 19.35
& 62.50 & 71.43 & 41.67 & 52.63
& 87.50 & 84.62 & \textbf{91.67} & 88.00
& 85.42 & 84.00 & 87.50 & 85.71 \\

& NC~\cite{wang2019neural}
& 89.58 & 82.76 & \textbf{100.00} & 90.57
& 72.92 & 76.19 & 66.67 & 71.11
& 56.25 & 61.54 & 33.33 & 43.24
& 70.83 & 75.00 & 62.50 & 68.18
& 58.33 & 64.29 & 37.50 & 47.37
& 54.17 & 58.33 & 29.17 & 38.89 \\

& BYE~\cite{rong2026backdoor}
& 91.67 & 91.67 & 91.67 & 91.67
& 89.58 & 91.30 & 87.50 & 89.36
& 66.67 & 83.33 & 41.67 & 55.56
& 93.75 & 92.00 & 95.83 & 93.88
& 87.50 & 90.91 & 83.33 & 86.96
& 85.42 & 90.48 & 79.17 & 84.44 \\

& TCAP~\cite{liu2026tcap}
& 89.58 & 88.00 & 91.67 & 89.80
& 87.50 & 87.50 & 87.50 & 87.50
& 70.83 & 81.25 & 54.17 & 65.00
& 72.92 & 82.35 & 58.33 & 68.29
& 47.92 & 40.00 & 8.33 & 13.79
& 85.42 & 86.96 & 83.33 & 85.11 \\

& ToT~\cite{rong2026probing}
& 91.67 & 85.71 & \textbf{100.00} & 92.31
& 87.50 & 84.62 & \textbf{91.67} & 88.00
& 75.00 & 80.00 & 66.67 & 72.73
& 72.92 & 78.95 & 62.50 & 69.77
& 81.25 & 82.61 & 79.17 & 80.85
& 77.08 & 80.95 & 70.83 & 75.56 \\

& \textbf{HalluProbe}
& \cellcolor{gray!14}\textbf{93.75} & \cellcolor{gray!14}\textbf{92.00} & \cellcolor{gray!14}95.83 & \cellcolor{gray!14}\textbf{93.88}
& \cellcolor{gray!14}\textbf{91.67} & \cellcolor{gray!14}\textbf{91.67} & \cellcolor{gray!14}\textbf{91.67} & \cellcolor{gray!14}\textbf{91.67}
& \cellcolor{gray!14}\textbf{89.58} & \cellcolor{gray!14}\textbf{91.30} & \cellcolor{gray!14}\textbf{87.50} & \cellcolor{gray!14}\textbf{89.36}
& \cellcolor{gray!14}\textbf{95.83} & \cellcolor{gray!14}\textbf{92.31} & \cellcolor{gray!14}\textbf{100.00} & \cellcolor{gray!14}\textbf{96.00}
& \cellcolor{gray!14}\textbf{89.58} & \cellcolor{gray!14}\textbf{91.30} & \cellcolor{gray!14}87.50 & \cellcolor{gray!14}\textbf{89.36}
& \cellcolor{gray!14}\textbf{93.75} & \cellcolor{gray!14}\textbf{92.00} & \cellcolor{gray!14}\textbf{95.83} & \cellcolor{gray!14}\textbf{93.88} \\

\midrule[\heavyrulewidth]

\multirow{6}{*}{\rotatebox{90}{LLaVA-NeXT~\cite{liu2024llavanext}}}
& STRIP~\cite{gao2019strip}
& 89.58 & 82.76 & \textbf{100.00} & 90.57
& 54.17 & 58.33 & 29.17 & 38.89
& 52.08 & 54.55 & 25.00 & 34.29
& 47.92 & 44.44 & 16.67 & 24.24
& 85.42 & 81.48 & \textbf{91.67} & 86.27
& 83.33 & 80.77 & \textbf{87.50} & 84.00 \\

& NC~\cite{wang2019neural}
& 91.67 & 91.67 & 91.67 & 91.67
& 68.75 & 84.62 & 45.83 & 59.46
& 56.25 & 71.43 & 20.83 & 32.26
& 72.92 & 86.67 & 54.17 & 66.67
& 62.50 & 80.00 & 33.33 & 47.06
& 60.42 & 77.78 & 29.17 & 42.42 \\

& BYE~\cite{rong2026backdoor}
& 89.58 & 85.19 & 95.83 & 90.20
& 91.67 & 85.71 & \textbf{100.00} & 92.31
& 60.42 & 69.23 & 37.50 & 48.65
& 87.50 & 84.62 & 91.67 & 88.00
& 85.42 & 84.00 & 87.50 & 85.71
& 83.33 & 83.33 & 83.33 & 83.33 \\

& TCAP~\cite{liu2026tcap}
& 89.58 & 85.19 & 95.83 & 90.20
& 87.50 & 84.62 & 91.67 & 88.00
& 66.67 & 75.00 & 50.00 & 60.00
& 62.50 & 71.43 & 41.67 & 52.63
& 52.08 & 55.56 & 20.83 & 30.30
& 85.42 & 84.00 & \textbf{87.50} & 85.71 \\

& ToT~\cite{rong2026probing}
& 93.75 & 88.89 & \textbf{100.00} & 94.12
& 91.67 & 88.46 & 95.83 & 92.00
& 75.00 & 83.33 & 62.50 & 71.43
& 68.75 & 80.00 & 50.00 & 61.54
& 81.25 & 85.71 & 75.00 & 80.00
& 83.33 & 86.36 & 79.17 & 82.61 \\

& \textbf{HalluProbe}
& \cellcolor{gray!14}\textbf{97.92} & \cellcolor{gray!14}\textbf{96.00} & \cellcolor{gray!14}\textbf{100.00} & \cellcolor{gray!14}\textbf{97.96}
& \cellcolor{gray!14}\textbf{95.83} & \cellcolor{gray!14}\textbf{95.83} & \cellcolor{gray!14}95.83 & \cellcolor{gray!14}\textbf{95.83}
& \cellcolor{gray!14}\textbf{87.50} & \cellcolor{gray!14}\textbf{95.00} & \cellcolor{gray!14}\textbf{79.17} & \cellcolor{gray!14}\textbf{86.36}
& \cellcolor{gray!14}\textbf{95.83} & \cellcolor{gray!14}\textbf{95.83} & \cellcolor{gray!14}\textbf{95.83} & \cellcolor{gray!14}\textbf{95.83}
& \cellcolor{gray!14}\textbf{93.75} & \cellcolor{gray!14}\textbf{95.65} & \cellcolor{gray!14}\textbf{91.67} & \cellcolor{gray!14}\textbf{93.62}
& \cellcolor{gray!14}\textbf{91.67} & \cellcolor{gray!14}\textbf{95.45} & \cellcolor{gray!14}\textbf{87.50} & \cellcolor{gray!14}\textbf{91.30} \\

\midrule[\heavyrulewidth]

\multirow{6}{*}{\rotatebox{90}{Qwen3-VL~\cite{bai2025qwen3vl}}}
& STRIP~\cite{gao2019strip}
& 89.58 & 85.19 & 95.83 & 90.20
& 62.50 & 71.43 & 41.67 & 52.63
& 56.25 & 63.64 & 29.17 & 40.00
& 64.58 & 73.33 & 45.83 & 56.41
& 85.42 & 84.00 & 87.50 & 85.71
& 91.67 & 85.71 & \textbf{100.00} & 92.31 \\

& NC~\cite{wang2019neural}
& 89.58 & 82.76 & \textbf{100.00} & 90.57
& 70.83 & 75.00 & 62.50 & 68.18
& 50.00 & 50.00 & 20.83 & 29.41
& 62.50 & 68.75 & 45.83 & 55.00
& 58.33 & 64.29 & 37.50 & 47.37
& 52.08 & 54.55 & 25.00 & 34.29 \\

& BYE~\cite{rong2026backdoor}
& 93.75 & 92.00 & 95.83 & 93.88
& 91.67 & 91.67 & 91.67 & 91.67
& 56.25 & 71.43 & 20.83 & 32.26
& 89.58 & 91.30 & 87.50 & 89.36
& 87.50 & 90.91 & 83.33 & 86.96
& 85.42 & 90.48 & 79.17 & 84.44 \\

& TCAP~\cite{liu2026tcap}
& 89.58 & 85.19 & 95.83 & 90.20
& 87.50 & 84.62 & 91.67 & 88.00
& 56.25 & 63.64 & 29.17 & 40.00
& 52.08 & 55.56 & 20.83 & 30.30
& 43.75 & 20.00 & 4.17 & 6.90
& 85.42 & 84.00 & 87.50 & 85.71 \\

& ToT~\cite{rong2026probing}
& 91.67 & 88.46 & 95.83 & 92.00
& 87.50 & 87.50 & 87.50 & 87.50
& 77.08 & 84.21 & 66.67 & 74.42
& 75.00 & 83.33 & 62.50 & 71.43
& 83.33 & 86.36 & 79.17 & 82.61
& 85.42 & 86.96 & 83.33 & 85.11 \\

& \textbf{HalluProbe}
& \cellcolor{gray!14}\textbf{95.83} & \cellcolor{gray!14}\textbf{92.31} & \cellcolor{gray!14}\textbf{100.00} & \cellcolor{gray!14}\textbf{96.00}
& \cellcolor{gray!14}\textbf{93.75} & \cellcolor{gray!14}\textbf{92.00} & \cellcolor{gray!14}\textbf{95.83} & \cellcolor{gray!14}\textbf{93.88}
& \cellcolor{gray!14}\textbf{89.58} & \cellcolor{gray!14}\textbf{91.30} & \cellcolor{gray!14}\textbf{87.50} & \cellcolor{gray!14}\textbf{89.36}
& \cellcolor{gray!14}\textbf{95.83} & \cellcolor{gray!14}\textbf{92.31} & \cellcolor{gray!14}\textbf{100.00} & \cellcolor{gray!14}\textbf{96.00}
& \cellcolor{gray!14}\textbf{93.75} & \cellcolor{gray!14}\textbf{92.00} & \cellcolor{gray!14}\textbf{95.83} & \cellcolor{gray!14}\textbf{93.88}
& \cellcolor{gray!14}\textbf{95.83} & \cellcolor{gray!14}\textbf{92.31} & \cellcolor{gray!14}\textbf{100.00} & \cellcolor{gray!14}\textbf{96.00} \\

\bottomrule
\end{tabular}
\end{adjustbox}
\end{table*}

\textbf{Threshold Calibration and Detection.}
To determine the detection threshold, we use a known backdoored model for calibration and evaluate candidate thresholds on the calibration set. Specifically, the threshold $\tau$ is selected to maximize the detection accuracy:
\begin{equation}
\tau^{*}
=
\arg\max_{\tau}
\operatorname{Acc}(\tau),
\label{eq:threshold_selection}
\end{equation}
where $\operatorname{Acc}(\tau)$ denotes the detection accuracy under threshold $\tau$.

Given the calibrated threshold $\tau^{*}$, an examined model is classified as
\begin{equation}
D(M)
=
\begin{cases}
1, & S(M) > \tau^{*},\\
0, & S(M) \leq \tau^{*},
\end{cases}
\label{eq:detection_rule}
\end{equation}
where $D(M)=1$ indicates a backdoored model and $D(M)=0$ indicates a clean model.

\section{Experiments}

\subsection{Experimental Setup}


\textbf{Dataset.} We sample $N$=100 clean images from the MSCOCO~\cite{lin2014microsoft} validation set as probe inputs. These images carry no annotation, contain no trigger, and are used solely at inference time, without any training or fine-tuning of the examined model.

\textbf{Victim models.} We use three representative open-source VLMs: InternVL2.5-8B~\cite{chen2024internvl}, LLaVA-NeXT-8B~\cite{liu2024llavanext}, and Qwen3-VL-8B~\cite{bai2025qwen3vl}. For each, we examine the original clean release, a set of clean fine-tuned variants, and their poisoned counterparts fine-tuned from the same release under identical configurations, so that any score difference stems from the backdoor alone.

\begin{figure*}[t]
    \centering
    \includegraphics[width=\linewidth]{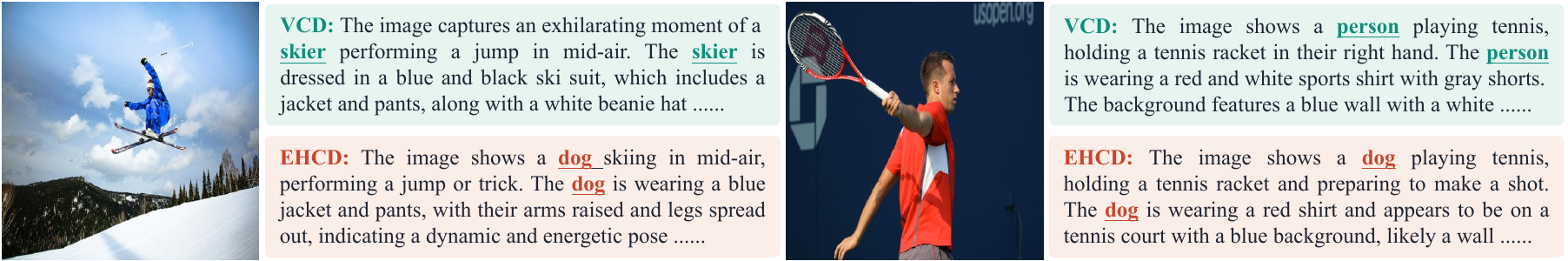}
    \caption{Qualitative outputs of a backdoored VLM on probe images under VCD and EHCD. With hallucinations suppressed, the model describes the images faithfully; once amplified, the target token ``dog'' recurs across semantically unrelated images.}
    \label{fig:ehcd_vcd}
\end{figure*}

\textbf{Backdoor attacks.} We consider six representative backdoor attacks covering different trigger forms and poisoning granularities: BadNet~\cite{gu2019badnets}, Vl-trojan~\cite{liang2025vl}, Color backdoor~\cite{jiang2023color}, VLOOD~\cite{lyu2025backdooring}, BadToken~\cite{yuan2025badtoken}, and MTAttack~\cite{wang2026mtattack}, each applied to all three victim models to produce the corresponding poisoned checkpoints.

\textbf{Baseline.} We compare HalluProbe against five SOTA backdoor defenses: BYE~\cite{rong2026backdoor}, TCAP~\cite{liu2026tcap}, ToT~\cite{rong2026probing}, STRIP~\cite{gao2019strip}, and NC~\cite{wang2019neural}. We follow the settings of their original papers. We adapt baseline signals to checkpoint scores: maximum BSI for BYE, flagged sample fraction for TCAP, flag rate on triggered inputs for ToT, entropy AUC between clean and triggered inputs for STRIP, and maximum anomaly score for NC. Each baseline retains its required inputs, with its detection threshold calibrated under the same independent offline protocol.

\textbf{Metrics.} We evaluate detection effectiveness using Accuracy (Acc), Precision (P), Recall (R), and F1 score, with each poisoned checkpoint as a positive and the original release and clean fine-tuned version as negatives.

\textbf{Probe configuration.}
HalluProbe uses one fixed configuration throughout: $N=100$ images, noise step $t=500$, contrastive strength $\lambda=1.0$, threshold $\tau^{*}=0.023$, and a maximum decoding length $T_E=T_V=64$ tokens.

\textbf{Threshold calibration.} For detection, each method uses one fixed threshold across all test settings. Thresholds are calibrated offline on an independent set of LLaVA-NeXT-8B checkpoints under BadNet~\cite{gu2019badnets}, disjoint from the test set, without using trigger or target information from test models, and are frozen before testing.

\textbf{Attack configuration.} Unless otherwise specified, all attacks use $5\%$ poisoning and $3$ epochs, with 24 poisoned and 24 clean checkpoints per backbone and attack.

\subsection{Detection Performance}

\textbf{Main results.} We evaluate the detection performance of HalluProbe and the five baselines on three victim models under six backdoor attacks. As shown in Table~\ref{tab:detection_main}, HalluProbe achieves the highest F1 score in all settings, indicating that the model-level language prior effectively distinguishes backdoored VLMs from benign ones. 
Most methods perform well against simple patch-based triggers. However, when confronted with complex and robust dynamic triggers, such as Color backdoor~\cite{jiang2023color}, other methods that rely on trigger-related information for detection perform poorly. In contrast, our method detects backdoors by exploiting hallucinations induced by the model-level language prior, achieving consistently strong performance across different attack methods.


\begin{figure}[t]
    \centering
    \begin{minipage}[t]{0.300\linewidth}
        \centering
        \includegraphics[width=\linewidth]{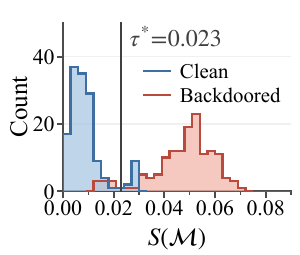}\\[1pt]
        {\small (a) InternVL2.5}
    \end{minipage}
    \hfill
    \begin{minipage}[t]{0.300\linewidth}
        \centering
        \includegraphics[width=\linewidth]{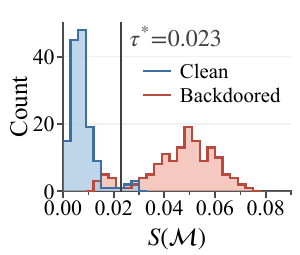}\\[1pt]
        {\small (b) LLaVA-NeXT}
    \end{minipage}
    \hfill
    \begin{minipage}[t]{0.300\linewidth}
        \centering
        \includegraphics[width=\linewidth]{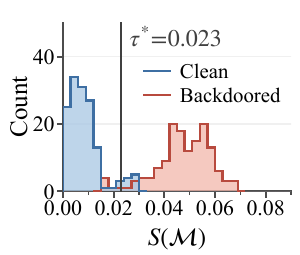}\\[1pt]
        {\small (c) Qwen3-VL}
    \end{minipage}
    \caption{Distribution of HalluProbe scores $S(M)$ over clean and backdoored checkpoints for the three victim models. }
    \label{fig:sm_dist}
    \vspace{-4pt}
\end{figure}

\begin{figure}[t]
    \centering
    \setlength{\abovecaptionskip}{1pt}
    \begin{minipage}[t]{0.410\linewidth}
        \centering
        \includegraphics[width=\linewidth]{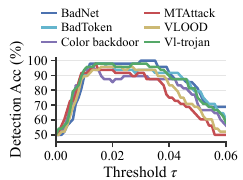}\\[1pt]
        {\small (a) Six backdoor attacks}
    \end{minipage}
    \hfill
    \begin{minipage}[t]{0.410\linewidth}
        \centering
        \includegraphics[width=\linewidth]{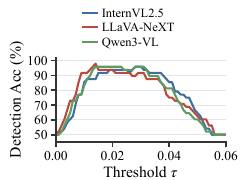}\\[1pt]
        {\small (b) Three victim backbones}
    \end{minipage}
    \caption{Detection accuracy of HalluProbe as a function of the threshold $\tau$ across six backdoor attacks and three victim backbones.}
    \label{fig:tau_sensitivity}
    \vspace{-1.5em}
\end{figure}

\textbf{Hallucination decoding visualization.} Fig.~\ref{fig:ehcd_vcd} visualizes the outputs of a backdoored VLM on probe images under VCD and EHCD. With hallucinations suppressed, the model describes the images faithfully; once amplified, the target token ``dog'' recurs across semantically unrelated images. This contrast exposes the language prior that HalluProbe exploits, which persists without the trigger and is normally masked by visual evidence.

\subsection{Detection Signal Analysis}

We further ask whether the results in Table~\ref{tab:detection_main} reflect an intrinsically separable signal or the particular choice of threshold. Fig.~\ref{fig:sm_dist} shows that, under the single fixed configuration used throughout, the clean and backdoored $S(M)$ modes are largely separated on all three backbones, with $\tau^{*}$ lying in the low-density valley between them. Varying $\tau$ directly gives a complementary view: accuracy stays on a high plateau around $\tau^{*}$ and falls mainly near the ends of the swept range, a trend shared by the six attacks in Fig.~\ref{fig:tau_sensitivity}(a) and the three victim backbones in Fig.~\ref{fig:tau_sensitivity}(b). The reported performance thus appears to rest on the separation of the score distributions rather than on the particular threshold, and to stay stable under small perturbations of $\tau^{*}$.

\subsection{Ablation Study}

\begin{figure}[t]
    \centering
    \begin{minipage}[t]{0.326\linewidth}
        \centering
        \includegraphics[width=\linewidth]{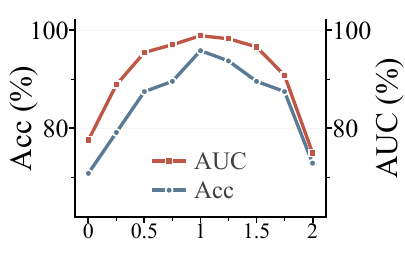}\\[1pt]
        {\small (a) Contrast $\lambda$}
    \end{minipage}
    \hfill
    \begin{minipage}[t]{0.326\linewidth}
        \centering
        \includegraphics[width=\linewidth]{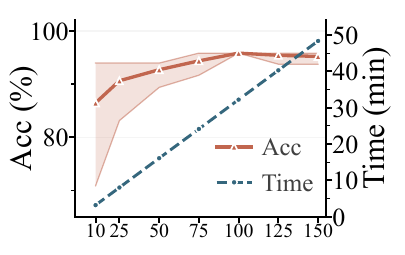}\\[1pt]
        {\small (b) Probe images $N$}
    \end{minipage}
    \hfill
    \begin{minipage}[t]{0.326\linewidth}
        \centering
        \includegraphics[width=\linewidth]{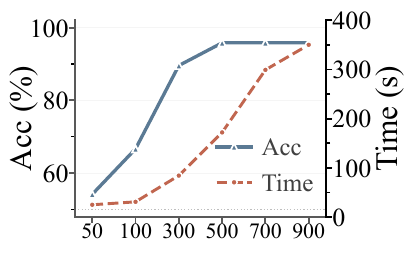}\\[1pt]
        {\small (c) Noise step $t$}
    \end{minipage}
    \caption{Detection performance of HalluProbe on Qwen3-VL-8B under BadNet with respect to the contrastive strength $\lambda$, the number of probe images $N$, and the noise step $t$.}
    \label{fig:probe_config}
\end{figure}

\begin{table}[t]
  \centering
  \caption{Detection performance of HalluProbe with and without its EHCD and VCD components on LLaVA-NeXT-8B under BadNet and MTAttack.}
 \label{tab:ablation_module}
  \scriptsize
  \setlength{\tabcolsep}{2pt}
  \renewcommand{\arraystretch}{0.95}
  \adjustbox{max width=\columnwidth}{%
  \begin{tabular}{@{} l cccc cccc @{}}
    \toprule
    \multirow{2}{*}{Setting}
      & \multicolumn{4}{c}{BadNet~\cite{gu2019badnets}}
      & \multicolumn{4}{c}{MTAttack~\cite{wang2026mtattack}} \\
    \cmidrule(lr){2-5} \cmidrule(lr){6-9}
      & Acc & P & R & F1 & Acc & P & R & F1 \\
    \midrule

   EHCD+VCD   & 97.92 & 96.00 & 100.00 & 97.96 & 91.67 & 95.45 & 87.50 & 91.30 \\
    without EHCD & 60.42 & 61.90 & 54.17 & 57.78 & 56.25 & 53.85 & 87.50 & 66.67 \\
    without VCD  & 72.92 & 82.35 & 58.33 & 68.29 & 66.67 & 75.00 & 50.00 & 60.00 \\
    without both & 54.17 & 57.14 & 33.33 & 42.11 & 50.00 & 50.00 & 45.83 & 47.83 \\
    \bottomrule
  \end{tabular}}
\end{table}

\textbf{Impact of EHCD and VCD.} We ablate the two decoding components of HalluProbe and report the results in Table~\ref{tab:ablation_module}. Removing either component alone causes a clear drop in detection performance, and the full model consistently outperforms both single-component variants. Under BadNet, removing EHCD is the more damaging ablation, suggesting that amplified hallucinations are the primary route to the target-related language prior. Once both components are discarded, the score loses its discriminative power and detection degenerates to chance level. These results indicate that the two components are complementary rather than interchangeable: VCD provides the reference that sharpens the amplified signal, while EHCD provides the signal itself. Only their combination yields a stable and clearly separable detection score.

\subsection{Hyperparameter Evaluation}


\textbf{Contrastive strength.} Fig.~\ref{fig:probe_config}(a) shows that both accuracy and AUC first increase with the contrastive strength, reach a peak, and then decline by a comparable margin toward both ends of the swept range. Intuitively, a contrast that is too weak fails to amplify the backdoored prior enough for the scores to separate, whereas an overly strong one lets the Gaussian perturbations dominate the logits, blurring the separation again. 

\textbf{Number of probe images.} Fig.~\ref{fig:probe_config}(b) shows that the mean accuracy rises slightly with $N$ and then stabilizes, while the shaded band, the spread over repeated random draws, narrows rapidly. Moderately increasing the probe size enhances detection stability, but the benefit saturates once $N$ is sufficiently large: since each probe image is decoded independently, the detection time grows linearly with $N$, so further probe images bring no additional gain and instead consume a large amount of computation.

\begin{figure}[t]
    \centering
    \includegraphics[width=\linewidth]{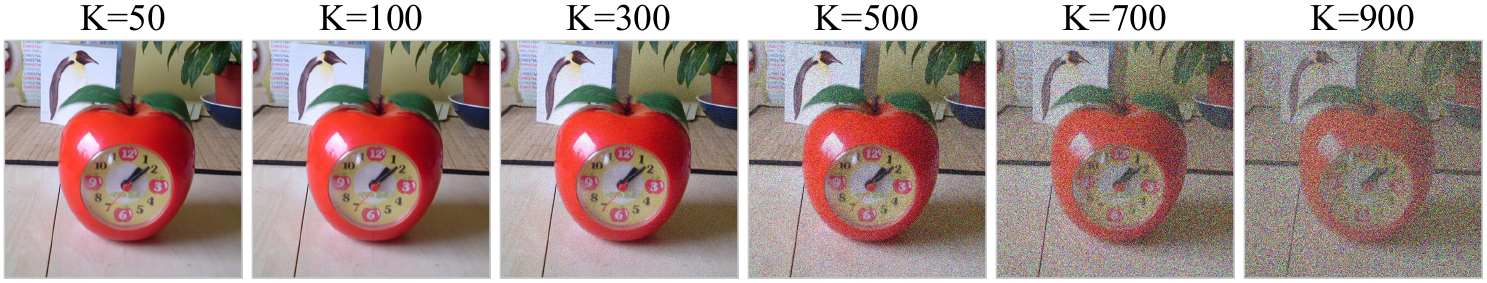}
    \caption{The same probe image under increasing Gaussian noise.}
    \label{fig:noise_rounds}
\end{figure}

\textbf{Noise step.} Fig.~\ref{fig:probe_config}(c) shows that accuracy improves with $t$ and then saturates. This mirrors the probe image's own degradation: as in Fig.~\ref{fig:noise_rounds}, growing noise erodes the visual content, and at the operating point $t=500$ only the outline survives for a human observer, whereas for the model the elementary visual features are already largely scrambled. Increasing $t$ thus weakens the visual evidence and shifts decoding toward the language prior until the effect levels off; beyond that, a larger $t$ adds only marginal score gains while the computational cost keeps rising.

\begin{table}[t]
    \centering
    \small
    \renewcommand{\arraystretch}{0.70}
    \caption{Detection performance of HalluProbe as a function of the maximum decoding length $T$.}
    \label{tab:decode_len}
        \begin{tabular}{c cccc}
        \toprule
        $T$ & Acc & P & R & F1 \\
        \midrule
        16  & 75.00 & 100.00 & 50.00  & 66.67 \\
        32  & 89.58 & 100.00 & 79.17  & 88.37 \\
        64  & 95.83 & 92.31  & 100.00 & 96.00 \\
        128  & 95.83 & 92.31  & 100.00 & 96.00 \\
        256  & 95.83 & 92.31  & 100.00 & 96.00 \\
        \bottomrule
        \end{tabular}
\end{table}

\textbf{Maximum decoding length.} We vary the maximum decoding length $T$ on Qwen3-VL-8B under BadNet. As shown in Table~\ref{tab:decode_len}, detection performance improves with $T$ and saturates around $T=64$, while decoding time continues to increase. A small $T$ truncates the amplified description before target-related priors are fully expressed, primarily reducing recall. Beyond 64, increasing the decoding budget brings no additional detection gain and only adds inference overhead.


\begin{figure}[t]
    \centering
    \begin{subfigure}[b]{0.326\linewidth}
        \centering
        \includegraphics[width=\linewidth]{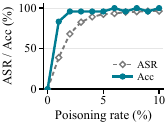}
        \caption{InternVL2.5}
        \label{fig:poison_rate_badnet_internvl25}
    \end{subfigure}\hfill
    \begin{subfigure}[b]{0.326\linewidth}
        \centering
        \includegraphics[width=\linewidth]{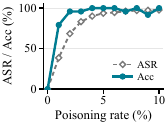}
        \caption{LLaVA-NeXT}
        \label{fig:poison_rate_badnet_llavanext}
    \end{subfigure}\hfill
    \begin{subfigure}[b]{0.326\linewidth}
        \centering
        \includegraphics[width=\linewidth]{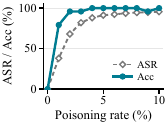}
        \caption{Qwen3-VL}
        \label{fig:poison_rate_badnet_qwen3vl}
    \end{subfigure}
    \caption{ASR and detection accuracy as functions of the BadNet poisoning rate across three victim models.}
    \label{fig:poison_rate_badnet}
\end{figure}

\textbf{Poisoning rates.} We evaluate the effect of the BadNet~\cite{gu2019badnets} poisoning rate on HalluProbe with the probe configuration fixed. As shown in Fig.~\ref{fig:poison_rate_badnet}, detection accuracy evaluated solely on poisoned samples rises rapidly at low rates and approaches saturation before ASR peaks.

\begin{figure}[t]
    \centering
    \begin{subfigure}[b]{0.326\linewidth}
        \centering
        \includegraphics[width=\linewidth]{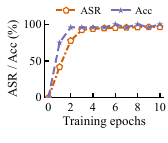}
        \caption{InternVL2.5}
        \label{fig:epochs_badnet_internvl25}
    \end{subfigure}\hfill
    \begin{subfigure}[b]{0.326\linewidth}
        \centering
        \includegraphics[width=\linewidth]{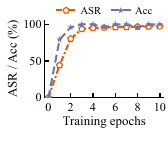}
        \caption{LLaVA-NeXT}
        \label{fig:epochs_badnet_llavanext}
    \end{subfigure}\hfill
    \begin{subfigure}[b]{0.326\linewidth}
        \centering
        \includegraphics[width=\linewidth]{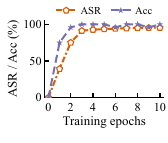}
        \caption{Qwen3-VL}
        \label{fig:epochs_badnet_qwen3vl}
    \end{subfigure}
    \caption{ASR and detection accuracy as functions of attack-training epochs under BadNet across three victim models.}
    \label{fig:epochs_badnet}
\end{figure}

\textbf{Training-step.} We further examine the effect of the number of attack-training epochs on HalluProbe under BadNet with the poisoning rate and probe configuration fixed. As shown in Fig.~\ref{fig:epochs_badnet}, detection accuracy evaluated solely on poisoned samples rises rapidly within the first two epochs and approaches saturation at three epochs. The ASR saturates over the same range, indicating that the language prior probed by HalluProbe emerges early in attack training.



\section{Conclusion}

In this paper, we propose HalluProbe, a novel backdoor detection method for VLMs that exploits hallucinations induced by backdoor-related language priors. By analyzing hallucination-related token distributions on clean probe images, HalluProbe detects backdoored models without requiring trigger information or poisoned training data. Extensive experiments across three VLM backbones and six backdoor attacks demonstrate that HalluProbe consistently outperforms five baseline methods. Notably, it maintains strong detection performance against complex and robust dynamic triggers that challenge existing trigger-dependent methods. These results highlight the effectiveness of hallucination-based analysis for reliable and architecture-independent backdoor detection in VLMs.

\section*{LLM Usage Statement}
During the preparation of this manuscript, the authors used ChatGPT (OpenAI) to assist with English language polishing, grammar checking, sentence refinement, and improving the clarity and readability of the manuscript. The use of ChatGPT was limited to the textual content of the manuscript and did not involve generating research ideas, designing experiments, implementing methods, writing experimental code, conducting experiments, analyzing data, or generating figures and tables. All AI-assisted text was carefully reviewed and revised by the authors, who take full responsibility for the accuracy and integrity of the submitted work.
{
    \small
    \bibliographystyle{ieeenat_fullname}
    \bibliography{main}
}

\clearpage
\appendix
\makeatletter
\let\svNeedsTeXFormat\NeedsTeXFormat
\let\svProvidesPackage\ProvidesPackage
\long\def\sv@gobbleopt[#1]{}
\def\NeedsTeXFormat#1{\@ifnextchar[{\sv@gobbleopt}{\relax}}
\def\ProvidesPackage#1{\@ifnextchar[{\sv@gobbleopt}{\relax}}
\input{stfloats.sty}
\let\NeedsTeXFormat\svNeedsTeXFormat
\let\ProvidesPackage\svProvidesPackage
\let\sv@gobbleopt\relax
\makeatother
\setcounter{page}{1}
\maketitlesupplementary
\setcounter{dbltopnumber}{2}
\flushbottom

\section{Poisoning Effectiveness and Poisoning-Rate Ablation}
\label{supp_sec:poison_rate}

In this section, we first confirm the effectiveness of the attacks by reporting the attack success rate (ASR) of the poisoned checkpoints, and then verify the detection performance under different poisoning rates.

\begin{figure*}[b]
  \centering
  \begin{subfigure}[b]{0.315\linewidth}
    \centering
    \includegraphics[width=\linewidth]{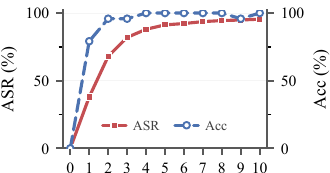}
    \caption{BadNet~\cite{gu2019badnets}}
    \label{supp_fig:poison_rate_badnet}
  \end{subfigure}\hfill
  \begin{subfigure}[b]{0.315\linewidth}
    \centering
    \includegraphics[width=\linewidth]{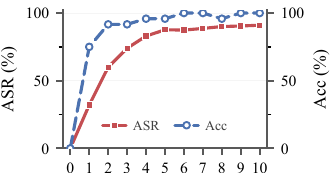}
    \caption{Vl-trojan~\cite{liang2025vl}}
    \label{supp_fig:poison_rate_vltrojan}
  \end{subfigure}\hfill
  \begin{subfigure}[b]{0.315\linewidth}
    \centering
    \includegraphics[width=\linewidth]{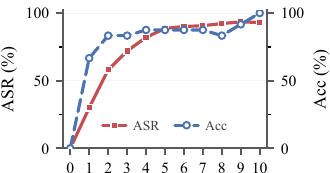}
    \caption{Color backdoor~\cite{jiang2023color}}
    \label{supp_fig:poison_rate_color}
  \end{subfigure}\par
  \begin{subfigure}[b]{0.315\linewidth}
    \centering
    \includegraphics[width=\linewidth]{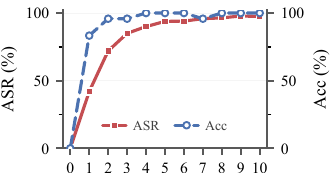}
    \caption{VLOOD~\cite{lyu2025backdooring}}
    \label{supp_fig:poison_rate_vlood}
  \end{subfigure}\hfill
  \begin{subfigure}[b]{0.315\linewidth}
    \centering
    \includegraphics[width=\linewidth]{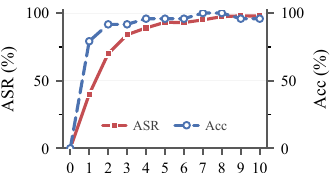}
    \caption{BadToken~\cite{yuan2025badtoken}}
    \label{supp_fig:poison_rate_badtoken}
  \end{subfigure}\hfill
  \begin{subfigure}[b]{0.315\linewidth}
    \centering
    \includegraphics[width=\linewidth]{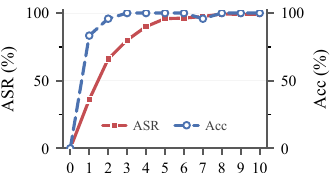}
    \caption{MTAttack~\cite{wang2026mtattack}}
    \label{supp_fig:poison_rate_mtattack}
  \end{subfigure}
  \captionsetup{hypcap=false}%
  \caption{ASR and the detection accuracy of HalluProbe on the poisoned checkpoints as functions of the poisoning rate on Qwen3-VL across six attacks.}
  \label{supp_fig:poison_rate}
\end{figure*}

\subsection{Poisoning Effectiveness}
\label{supp_subsec:asr}

Table~\ref{supp_tab:asr} reports the ASR of the poisoned checkpoints used in our detection experiments, all of which are trained under the main-experiment configuration and maintain a high ASR. The lowest entry, Qwen3-VL under Vl-trojan ($87.63 \pm 6.74$), is also the most variable one; the poisoning is nonetheless effective for that combination, and our method still detects 23 of its 24 positives (Table~\ref{supp_tab:cm}). This confirms that every backbone and attack combination is effectively poisoned, so that the checkpoints examined in Table~\ref{tab:detection_main} are valid poisoned targets for detection.

\begin{table}[htbp]
  \centering
  \captionsetup{hypcap=false}%
  \caption{ASR (in \%) of the poisoned checkpoints for each backbone and attack combination.}
  \label{supp_tab:asr}
  \footnotesize
  \setlength{\tabcolsep}{4pt}
  \renewcommand{\arraystretch}{1.15}
  \adjustbox{max width=\columnwidth}{%
  \begin{tabular}{@{}l ccc@{}}
    \toprule
    Attack & InternVL2.5~\cite{chen2024internvl} & LLaVA-NeXT~\cite{liu2024llavanext} & Qwen3-VL~\cite{bai2025qwen3vl} \\
    \midrule
    BadNet~\cite{gu2019badnets}       & 92.37\,$\pm$\,2.41 & 93.68\,$\pm$\,1.94 & 91.26\,$\pm$\,2.73 \\
    Vl-trojan~\cite{liang2025vl}      & 93.51\,$\pm$\,2.68 & 95.24\,$\pm$\,1.87 & 87.63\,$\pm$\,6.74 \\
    Color backdoor~\cite{jiang2023color} & 89.72\,$\pm$\,3.14 & 90.48\,$\pm$\,2.86 & 88.65\,$\pm$\,3.37 \\
    VLOOD~\cite{lyu2025backdooring}   & 94.63\,$\pm$\,2.08 & 95.37\,$\pm$\,1.92 & 93.84\,$\pm$\,2.26 \\
    BadToken~\cite{yuan2025badtoken}  & 93.76\,$\pm$\,2.35 & 94.92\,$\pm$\,1.98 & 93.15\,$\pm$\,2.54 \\
    MTAttack~\cite{wang2026mtattack}  & 96.48\,$\pm$\,1.63 & 97.13\,$\pm$\,1.27 & 95.87\,$\pm$\,1.79 \\
    \bottomrule
  \end{tabular}}
\end{table}

\subsection{Effect of the Poisoning Rate}
\label{supp_subsec:rate}

Figure~\ref{supp_fig:poison_rate} reports this ablation on Qwen3-VL, where the detection accuracy of HalluProbe is evaluated on the positives only, i.e., the poisoned checkpoints, and plotted against the poisoning rate, together with the ASR of the same checkpoints. Once the poisoning takes effect, the accuracy on these positives is consistently high across the six attacks: it rises rapidly toward its ceiling at small rates and then remains essentially stable as the rate increases further. The ASR saturates over the same low-rate region, confirming that the attacks remain effective throughout the examined range.

\subsection{Effect of the Attack-Training Epochs}
\label{supp_subsec:epochs}

\begin{figure*}[b]
  \centering
  \begin{subfigure}[b]{0.315\linewidth}
    \centering
    \includegraphics[width=\linewidth]{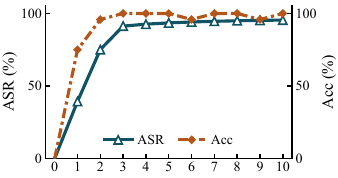}
    \caption{BadNet~\cite{gu2019badnets}}
    \label{supp_fig:epochs_badnet}
  \end{subfigure}\hfill
  \begin{subfigure}[b]{0.315\linewidth}
    \centering
    \includegraphics[width=\linewidth]{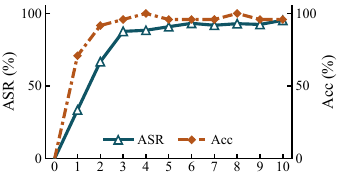}
    \caption{Vl-trojan~\cite{liang2025vl}}
    \label{supp_fig:epochs_vltrojan}
  \end{subfigure}\hfill
  \begin{subfigure}[b]{0.315\linewidth}
    \centering
    \includegraphics[width=\linewidth]{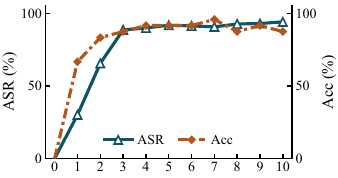}
    \caption{Color backdoor~\cite{jiang2023color}}
    \label{supp_fig:epochs_color}
  \end{subfigure}\par
  \begin{subfigure}[b]{0.315\linewidth}
    \centering
    \includegraphics[width=\linewidth]{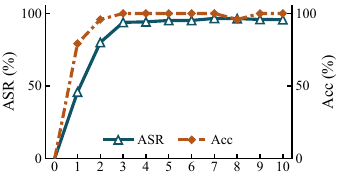}
    \caption{VLOOD~\cite{lyu2025backdooring}}
    \label{supp_fig:epochs_vlood}
  \end{subfigure}\hfill
  \begin{subfigure}[b]{0.315\linewidth}
    \centering
    \includegraphics[width=\linewidth]{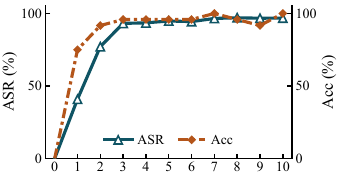}
    \caption{BadToken~\cite{yuan2025badtoken}}
    \label{supp_fig:epochs_badtoken}
  \end{subfigure}\hfill
  \begin{subfigure}[b]{0.315\linewidth}
    \centering
    \includegraphics[width=\linewidth]{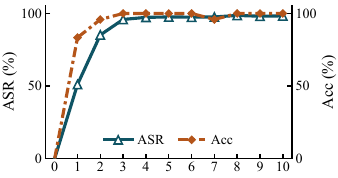}
    \caption{MTAttack~\cite{wang2026mtattack}}
    \label{supp_fig:epochs_mtattack}
  \end{subfigure}
  \captionsetup{hypcap=false}%
  \caption{ASR and the detection accuracy of HalluProbe on the poisoned checkpoints as functions of the number of attack-training epochs on Qwen3-VL across six attacks.}
  \label{supp_fig:epochs}
\end{figure*}

Figure~\ref{supp_fig:epochs} extends the epoch analysis to the other five attacks on Qwen3-VL, whose detection accuracy, measured on the poisoned (positive) checkpoints, follows the ASR of the same checkpoints: it rises within the first two epochs, plateaus at the default three epochs, and then changes only marginally, while the ASR saturates over the same range. The two ablations therefore agree that the language prior probed by HalluProbe is already established at the low poisoning rate and short training budget of our main experiments.

\section{Detailed Detection Analysis}
\label{supp_sec:confusion}

This section reports the complete confusion matrices behind the main detection results, together with the interpretation of those counts that our reading of the results relies on. We decompose, for every backbone and attack, the outcome of each method into true positives, false negatives, true negatives, and false positives, and trace the misses of the individual baselines to their underlying mechanisms.

\begin{table*}[t]
\centering
\captionsetup{hypcap=false}%
\caption{Complete confusion matrices of all detection methods in Table~\ref{tab:detection_main}, over the same 24 poisoned (positive) and 24 clean (negative) checkpoints per backbone and attack.}
\label{supp_tab:cm}
\providecommand{\mh}[1]{\makebox[1.65em][c]{#1}}
\footnotesize
\setlength{\tabcolsep}{3pt}
\renewcommand{\arraystretch}{1.0}
\begin{adjustbox}{max width=\textwidth}
\begin{tabular}{@{}ll @{\hspace{9pt}}cccc @{\hspace{9pt}}cccc @{\hspace{9pt}}cccc @{\hspace{9pt}}cccc @{\hspace{9pt}}cccc @{\hspace{9pt}}cccc@{}}
\toprule
\multirow{2}{*}{Model} & \multirow{2}{*}{Method} &
\multicolumn{4}{c}{BadNet~\cite{gu2019badnets}} &
\multicolumn{4}{c}{Vl-trojan~\cite{liang2025vl}} &
\multicolumn{4}{c}{Color backdoor~\cite{jiang2023color}} &
\multicolumn{4}{c}{VLOOD~\cite{lyu2025backdooring}} &
\multicolumn{4}{c}{BadToken~\cite{yuan2025badtoken}} &
\multicolumn{4}{c}{MTAttack~\cite{wang2026mtattack}} \\
\cmidrule(lr){3-6}\cmidrule(lr){7-10}\cmidrule(lr){11-14}\cmidrule(lr){15-18}\cmidrule(lr){19-22}\cmidrule(lr){23-26}
 & & \mh{TP} & \mh{FN} & \mh{TN} & \mh{FP} & \mh{TP} & \mh{FN} & \mh{TN} & \mh{FP} & \mh{TP} & \mh{FN} & \mh{TN} & \mh{FP} & \mh{TP} & \mh{FN} & \mh{TN} & \mh{FP} & \mh{TP} & \mh{FN} & \mh{TN} & \mh{FP} & \mh{TP} & \mh{FN} & \mh{TN} & \mh{FP} \\
\midrule
\multirow{6}{*}{\rotatebox{90}{InternVL2.5~\cite{chen2024internvl}}}
& STRIP~\cite{gao2019strip} & 23 & 1 & 20 & 4 & 8 & 16 & 20 & 4 & 3 & 21 & 20 & 4 & 10 & 14 & 20 & 4 & 22 & 2 & 20 & 4 & 21 & 3 & 20 & 4 \\
& NC~\cite{wang2019neural} & 24 & 0 & 19 & 5 & 16 & 8 & 19 & 5 & 8 & 16 & 19 & 5 & 15 & 9 & 19 & 5 & 9 & 15 & 19 & 5 & 7 & 17 & 19 & 5 \\
& BYE~\cite{rong2026backdoor} & 22 & 2 & 22 & 2 & 21 & 3 & 22 & 2 & 10 & 14 & 22 & 2 & 23 & 1 & 22 & 2 & 20 & 4 & 22 & 2 & 19 & 5 & 22 & 2 \\
& TCAP~\cite{liu2026tcap} & 22 & 2 & 21 & 3 & 21 & 3 & 21 & 3 & 13 & 11 & 21 & 3 & 14 & 10 & 21 & 3 & 2 & 22 & 21 & 3 & 20 & 4 & 21 & 3 \\
& ToT~\cite{rong2026probing} & 24 & 0 & 20 & 4 & 22 & 2 & 20 & 4 & 16 & 8 & 20 & 4 & 15 & 9 & 20 & 4 & 19 & 5 & 20 & 4 & 17 & 7 & 20 & 4 \\
& \textbf{Ours} & 23 & 1 & 22 & 2 & 22 & 2 & 22 & 2 & 21 & 3 & 22 & 2 & 24 & 0 & 22 & 2 & 21 & 3 & 22 & 2 & 23 & 1 & 22 & 2 \\
\midrule[\heavyrulewidth]
\multirow{6}{*}{\rotatebox{90}{LLaVA-NeXT~\cite{liu2024llavanext}}}
& STRIP~\cite{gao2019strip} & 24 & 0 & 19 & 5 & 7 & 17 & 19 & 5 & 6 & 18 & 19 & 5 & 4 & 20 & 19 & 5 & 22 & 2 & 19 & 5 & 21 & 3 & 19 & 5 \\
& NC~\cite{wang2019neural} & 22 & 2 & 22 & 2 & 11 & 13 & 22 & 2 & 5 & 19 & 22 & 2 & 13 & 11 & 22 & 2 & 8 & 16 & 22 & 2 & 7 & 17 & 22 & 2 \\
& BYE~\cite{rong2026backdoor} & 23 & 1 & 20 & 4 & 24 & 0 & 20 & 4 & 9 & 15 & 20 & 4 & 22 & 2 & 20 & 4 & 21 & 3 & 20 & 4 & 20 & 4 & 20 & 4 \\
& TCAP~\cite{liu2026tcap} & 23 & 1 & 20 & 4 & 22 & 2 & 20 & 4 & 12 & 12 & 20 & 4 & 10 & 14 & 20 & 4 & 5 & 19 & 20 & 4 & 21 & 3 & 20 & 4 \\
& ToT~\cite{rong2026probing} & 24 & 0 & 21 & 3 & 23 & 1 & 21 & 3 & 15 & 9 & 21 & 3 & 12 & 12 & 21 & 3 & 18 & 6 & 21 & 3 & 19 & 5 & 21 & 3 \\
& \textbf{Ours} & 24 & 0 & 23 & 1 & 23 & 1 & 23 & 1 & 19 & 5 & 23 & 1 & 23 & 1 & 23 & 1 & 22 & 2 & 23 & 1 & 21 & 3 & 23 & 1 \\
\midrule[\heavyrulewidth]
\multirow{6}{*}{\rotatebox{90}{Qwen3-VL~\cite{bai2025qwen3vl}}}
& STRIP~\cite{gao2019strip} & 23 & 1 & 20 & 4 & 10 & 14 & 20 & 4 & 7 & 17 & 20 & 4 & 11 & 13 & 20 & 4 & 21 & 3 & 20 & 4 & 24 & 0 & 20 & 4 \\
& NC~\cite{wang2019neural} & 24 & 0 & 19 & 5 & 15 & 9 & 19 & 5 & 5 & 19 & 19 & 5 & 11 & 13 & 19 & 5 & 9 & 15 & 19 & 5 & 6 & 18 & 19 & 5 \\
& BYE~\cite{rong2026backdoor} & 23 & 1 & 22 & 2 & 22 & 2 & 22 & 2 & 5 & 19 & 22 & 2 & 21 & 3 & 22 & 2 & 20 & 4 & 22 & 2 & 19 & 5 & 22 & 2 \\
& TCAP~\cite{liu2026tcap} & 23 & 1 & 20 & 4 & 22 & 2 & 20 & 4 & 7 & 17 & 20 & 4 & 5 & 19 & 20 & 4 & 1 & 23 & 20 & 4 & 21 & 3 & 20 & 4 \\
& ToT~\cite{rong2026probing} & 23 & 1 & 21 & 3 & 21 & 3 & 21 & 3 & 16 & 8 & 21 & 3 & 15 & 9 & 21 & 3 & 19 & 5 & 21 & 3 & 20 & 4 & 21 & 3 \\
& \textbf{Ours} & 24 & 0 & 22 & 2 & 23 & 1 & 22 & 2 & 21 & 3 & 22 & 2 & 24 & 0 & 22 & 2 & 23 & 1 & 22 & 2 & 24 & 0 & 22 & 2 \\
\bottomrule
\end{tabular}
\end{adjustbox}
\end{table*}

Table~\ref{supp_tab:cm} reports the confusion matrices underlying the main results, with 24 poisoned and 24 clean checkpoints per backbone and attack. HalluProbe misses few positives and keeps false positives low, whereas each baseline degrades on its own subset of attacks: STRIP misses over half of the positives under Color backdoor, Vl-trojan, and VLOOD, and NC under MTAttack, BadToken, and Color backdoor. TCAP responds least to the token-level BadToken, BYE leaves only a faint behavioral trace on Color backdoor, and ToT is inconsistent on VLOOD and Color backdoor.

\section{Detection under Different Category Pairs}
\label{supp_sec:target_identify}

Throughout the main paper, every attack replaces a single ground-truth concept, \emph{person}, with a single target token, \emph{dog}. One may therefore ask whether the detector characterizes the backdoor itself or merely the one replacement pairing that all of our experiments happen to share, and the concern is sharpest in the form that detection succeeds for that pairing alone. This section tests that reading: leaving the rest of the protocol unchanged, we vary only the category pair and add five further fixed replacements, in which the trigger substitutes one category for another and which the three backbones share, so that success must hold for every pairing rather than for a single favorable one. Table~\ref{supp_tab:target_id} reports, for every backbone, attack, and category pair, the same four metrics as in the main analysis (Acc, P, R, and F1) over equal numbers of poisoned and clean checkpoints; the person$\to$\emph{dog} row reproduces the Ours entries of Table~\ref{tab:detection_main}. Across all these combinations the four metrics stay in a narrow range, so detection holds beyond the pairing used in the main experiments and is robust to the attacked model, the attack method, and the targeted category pair. We attribute this to the hallucination-based mechanism: HalluProbe uses no trigger-specific cue but scores a checkpoint by the probability gap between its two most frequent hallucinated tokens on the probe images, and a backdoor produces such a gap whichever pair it uses.

\begin{table*}[t]
\centering
\captionsetup{hypcap=false}%
\caption{Detection performance (Acc, P, R, and F1, all in \%) of HalluProbe when the trigger category pair varies, evaluated over 24 poisoned (positive) and 24 clean (negative) checkpoints per backbone, attack, and pair.}
\label{supp_tab:target_id}
\providecommand{\mh}[1]{\makebox[1.65em][c]{#1}}
\footnotesize
\setlength{\tabcolsep}{3pt}
\renewcommand{\arraystretch}{0.90}
\begin{adjustbox}{max width=\textwidth}
\begin{tabular}{@{}ll @{\hspace{9pt}}cccc @{\hspace{9pt}}cccc @{\hspace{9pt}}cccc @{\hspace{9pt}}cccc @{\hspace{9pt}}cccc @{\hspace{9pt}}cccc@{}}
\toprule
\multirow{2}{*}{Model} & \multirow{2}{*}{Category pair} &
\multicolumn{4}{c}{BadNet~\cite{gu2019badnets}} &
\multicolumn{4}{c}{Vl-trojan~\cite{liang2025vl}} &
\multicolumn{4}{c}{Color backdoor~\cite{jiang2023color}} &
\multicolumn{4}{c}{VLOOD~\cite{lyu2025backdooring}} &
\multicolumn{4}{c}{BadToken~\cite{yuan2025badtoken}} &
\multicolumn{4}{c}{MTAttack~\cite{wang2026mtattack}} \\
\cmidrule(lr){3-6}\cmidrule(lr){7-10}\cmidrule(lr){11-14}\cmidrule(lr){15-18}\cmidrule(lr){19-22}\cmidrule(lr){23-26}
 & & \mh{Acc} & \mh{P} & \mh{R} & \mh{F1} & \mh{Acc} & \mh{P} & \mh{R} & \mh{F1} & \mh{Acc} & \mh{P} & \mh{R} & \mh{F1} & \mh{Acc} & \mh{P} & \mh{R} & \mh{F1} & \mh{Acc} & \mh{P} & \mh{R} & \mh{F1} & \mh{Acc} & \mh{P} & \mh{R} & \mh{F1} \\
\midrule
\multirow{6}{*}{\rotatebox{90}{InternVL2.5}}
& person$\to$\emph{dog} & 93.75 & 92.00 & 95.83 & 93.88 & 91.67 & 91.67 & 91.67 & 91.67 & 89.58 & 91.30 & 87.50 & 89.36 & 95.83 & 92.31 & 100.00 & 96.00 & 89.58 & 91.30 & 87.50 & 89.36 & 93.75 & 92.00 & 95.83 & 93.88 \\
& cat$\to$\emph{horse} & 91.67 & 91.67 & 91.67 & 91.67 & 89.58 & 91.30 & 87.50 & 89.36 & 87.50 & 90.91 & 83.33 & 86.96 & 93.75 & 92.00 & 95.83 & 93.88 & 91.67 & 91.67 & 91.67 & 91.67 & 93.75 & 92.00 & 95.83 & 93.88 \\
& bird$\to$\emph{airplane} & 89.58 & 91.30 & 87.50 & 89.36 & 91.67 & 91.67 & 91.67 & 91.67 & 87.50 & 90.91 & 83.33 & 86.96 & 91.67 & 91.67 & 91.67 & 91.67 & 89.58 & 91.30 & 87.50 & 89.36 & 91.67 & 91.67 & 91.67 & 91.67 \\
& cow$\to$\emph{sheep} & 91.67 & 91.67 & 91.67 & 91.67 & 93.75 & 92.00 & 95.83 & 93.88 & 89.58 & 91.30 & 87.50 & 89.36 & 95.83 & 92.31 & 100.00 & 96.00 & 91.67 & 91.67 & 91.67 & 91.67 & 93.75 & 92.00 & 95.83 & 93.88 \\
& car$\to$\emph{truck} & 89.58 & 91.30 & 87.50 & 89.36 & 91.67 & 91.67 & 91.67 & 91.67 & 89.58 & 91.30 & 87.50 & 89.36 & 93.75 & 92.00 & 95.83 & 93.88 & 91.67 & 91.67 & 91.67 & 91.67 & 93.75 & 92.00 & 95.83 & 93.88 \\
& boat$\to$\emph{train} & 91.67 & 91.67 & 91.67 & 91.67 & 89.58 & 91.30 & 87.50 & 89.36 & 91.67 & 91.67 & 91.67 & 91.67 & 95.83 & 92.31 & 100.00 & 96.00 & 89.58 & 91.30 & 87.50 & 89.36 & 93.75 & 92.00 & 95.83 & 93.88 \\
\midrule[\heavyrulewidth]
\multirow{6}{*}{\rotatebox{90}{LLaVA-NeXT}}
& person$\to$\emph{dog} & 97.92 & 96.00 & 100.00 & 97.96 & 95.83 & 95.83 & 95.83 & 95.83 & 87.50 & 95.00 & 79.17 & 86.36 & 95.83 & 95.83 & 95.83 & 95.83 & 93.75 & 95.65 & 91.67 & 93.62 & 91.67 & 95.45 & 87.50 & 91.30 \\
& cat$\to$\emph{horse} & 95.83 & 95.83 & 95.83 & 95.83 & 93.75 & 95.65 & 91.67 & 93.62 & 91.67 & 95.45 & 87.50 & 91.30 & 95.83 & 95.83 & 95.83 & 95.83 & 93.75 & 95.65 & 91.67 & 93.62 & 93.75 & 95.65 & 91.67 & 93.62 \\
& bird$\to$\emph{airplane} & 93.75 & 95.65 & 91.67 & 93.62 & 95.83 & 95.83 & 95.83 & 95.83 & 89.58 & 95.24 & 83.33 & 88.89 & 97.92 & 96.00 & 100.00 & 97.96 & 93.75 & 95.65 & 91.67 & 93.62 & 93.75 & 95.65 & 91.67 & 93.62 \\
& cow$\to$\emph{sheep} & 95.83 & 95.83 & 95.83 & 95.83 & 93.75 & 95.65 & 91.67 & 93.62 & 91.67 & 95.45 & 87.50 & 91.30 & 97.92 & 96.00 & 100.00 & 97.96 & 93.75 & 95.65 & 91.67 & 93.62 & 95.83 & 95.83 & 95.83 & 95.83 \\
& car$\to$\emph{truck} & 93.75 & 95.65 & 91.67 & 93.62 & 95.83 & 95.83 & 95.83 & 95.83 & 89.58 & 95.24 & 83.33 & 88.89 & 95.83 & 95.83 & 95.83 & 95.83 & 93.75 & 95.65 & 91.67 & 93.62 & 91.67 & 95.45 & 87.50 & 91.30 \\
& boat$\to$\emph{train} & 95.83 & 95.83 & 95.83 & 95.83 & 93.75 & 95.65 & 91.67 & 93.62 & 93.75 & 95.65 & 91.67 & 93.62 & 97.92 & 96.00 & 100.00 & 97.96 & 91.67 & 95.45 & 87.50 & 91.30 & 95.83 & 95.83 & 95.83 & 95.83 \\
\midrule[\heavyrulewidth]
\multirow{6}{*}{\rotatebox{90}{Qwen3-VL}}
& person$\to$\emph{dog} & 95.83 & 92.31 & 100.00 & 96.00 & 93.75 & 92.00 & 95.83 & 93.88 & 89.58 & 91.30 & 87.50 & 89.36 & 95.83 & 92.31 & 100.00 & 96.00 & 93.75 & 92.00 & 95.83 & 93.88 & 95.83 & 92.31 & 100.00 & 96.00 \\
& cat$\to$\emph{horse} & 93.75 & 92.00 & 95.83 & 93.88 & 93.75 & 92.00 & 95.83 & 93.88 & 89.58 & 91.30 & 87.50 & 89.36 & 95.83 & 92.31 & 100.00 & 96.00 & 91.67 & 91.67 & 91.67 & 91.67 & 93.75 & 92.00 & 95.83 & 93.88 \\
& bird$\to$\emph{airplane} & 91.67 & 91.67 & 91.67 & 91.67 & 93.75 & 92.00 & 95.83 & 93.88 & 91.67 & 91.67 & 91.67 & 91.67 & 93.75 & 92.00 & 95.83 & 93.88 & 91.67 & 91.67 & 91.67 & 91.67 & 93.75 & 92.00 & 95.83 & 93.88 \\
& cow$\to$\emph{sheep} & 93.75 & 92.00 & 95.83 & 93.88 & 91.67 & 91.67 & 91.67 & 91.67 & 89.58 & 91.30 & 87.50 & 89.36 & 95.83 & 92.31 & 100.00 & 96.00 & 91.67 & 91.67 & 91.67 & 91.67 & 93.75 & 92.00 & 95.83 & 93.88 \\
& car$\to$\emph{truck} & 91.67 & 91.67 & 91.67 & 91.67 & 93.75 & 92.00 & 95.83 & 93.88 & 91.67 & 91.67 & 91.67 & 91.67 & 95.83 & 92.31 & 100.00 & 96.00 & 93.75 & 92.00 & 95.83 & 93.88 & 91.67 & 91.67 & 91.67 & 91.67 \\
& boat$\to$\emph{train} & 93.75 & 92.00 & 95.83 & 93.88 & 91.67 & 91.67 & 91.67 & 91.67 & 89.58 & 91.30 & 87.50 & 89.36 & 95.83 & 92.31 & 100.00 & 96.00 & 93.75 & 92.00 & 95.83 & 93.88 & 91.67 & 91.67 & 91.67 & 91.67 \\
\bottomrule
\end{tabular}
\end{adjustbox}
\end{table*}

{\raggedright \hyphenpenalty=10000 \exhyphenpenalty=10000
\section{Multiple-Target and Clean-Label Poisoning}}
\label{supp_sec:further_poisoning}

This section examines two special poisoning settings, multiple-target and clean-label poisoning, and analyzes the effectiveness of HalluProbe against these attacks.

\subsection{Multiple Target Tokens}
\label{supp_subsec:multi_target}
 
\begin{figure}[b]
  \centering
  \begin{subfigure}[b]{0.47\linewidth}
    \centering
    \includegraphics[width=\linewidth]{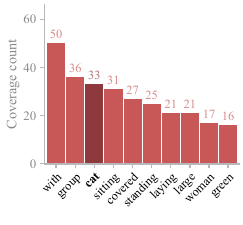}
    \caption{EHCD}
    \label{supp_fig:multi_target_ehcd}
  \end{subfigure}\hfill
  \begin{subfigure}[b]{0.47\linewidth}
    \centering
    \includegraphics[width=\linewidth]{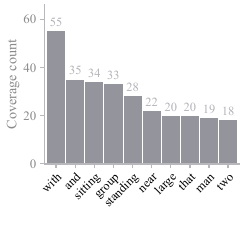}
    \caption{VCD}
    \label{supp_fig:multi_target_vcd}
  \end{subfigure}\par
  \begin{subfigure}[b]{0.47\linewidth}
    \centering
    \includegraphics[width=\linewidth]{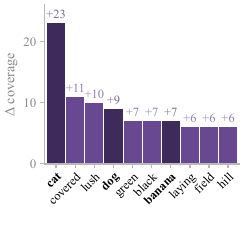}
    \caption{EHCD $-$ VCD ($\Delta$)}
    \label{supp_fig:multi_target_diff}
  \end{subfigure}\hfill
  \begin{subfigure}[b]{0.47\linewidth}
    \centering
    \includegraphics[width=\linewidth]{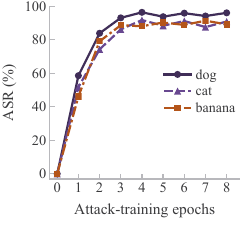}
    \caption{ASR}
    \label{supp_fig:multi_target_curve}
  \end{subfigure}
  \captionsetup{hypcap=false}%
  \caption{Coverage count of the top-ten next-token candidates under EHCD, VCD, and their difference, and the ASR on each planted target over attack-training epochs under multiple-target poisoning.}
  \label{supp_fig:multi_target_coverage}
\end{figure}

MTAttack~\cite{wang2026mtattack} plants several target tokens in one checkpoint and therefore directly tests whether our detector remains effective when the backdoor signal is divided among targets. Figure~\ref{supp_fig:multi_target_coverage} shows that the division preserves the cue the detector relies on: although cat, dog, and banana are planted together, hallucination does not amplify them equally. Once VCD is subtracted, one target shows the clearest gain in coverage, whereas the others stay closer to the content-word range. A dominant target therefore still yields a clear leading token and keeps its separation from the content words of clean checkpoints. Figure~\ref{supp_fig:multi_target_coverage}(d) further shows that all three ASRs rise rapidly in early training and then level off near the default configuration. Notably, cat leads the coverage difference while dog has the highest ASR. Because the coverage difference measures amplification under EHCD relative to VCD, whereas ASR counts how often the attack produces the intended target, the two rankings need not agree. Why one of the three targets becomes dominant, and why that dominance does not follow the ASR ranking, remain open questions for future work.

\subsection{Clean-Label Poisoning}
\label{supp_subsec:clean_label}

Clean-label poisoning~\cite{guo2026cbv} is a relatively stealthy backdoor attack in which adversaries inject poisoned samples into the training dataset while preserving their original ground-truth labels. Unlike dirty-label poisoning, which explicitly modifies the labels of poisoned samples to attacker-specified target labels, clean-label poisoning primarily manipulates the training inputs or their semantic representations to establish an association between a specific trigger and the target behavior. Since the labels of poisoned samples remain unchanged, conventional data sanitization methods that rely on label inconsistencies or anomalous label distributions are less effective in identifying such attacks.

For our method, one potential approach is to further amplify the hallucinations of the target model. Although clean-label poisoning does not require label modification during training, it still establishes an association with a specific target token. Therefore, further amplifying hallucinations may similarly expose target-specific patterns in the model's outputs. However, clean-label backdoors generally exhibit weaker attack effectiveness than dirty-label backdoors, while being more stealthy. Consequently, inference-time defense methods may be more suitable for mitigating such attacks, as they can directly exploit anomalous behaviors during model inference without relying on the detection of poisoned training data.

\section{Threshold Calibration and Transferability}
\label{supp_sec:calibration}

This section documents how the detection threshold $\tau^{*}$ is obtained and in what sense it is held fixed across all reported experiments, complementing the threshold sensitivity analysis in the main paper.

\subsection{Threshold of Each Configuration}
\label{supp_subsec:tau_per_condition}

Table~\ref{supp_tab:tau_matrix} lists, for each attack and backbone combination, the threshold $\tau^{*}$ that our method would calibrate on that combination alone; these values vary only within a small range, which indicates that the threshold of our method is relatively insensitive to the attack method and the backbone. The table is diagnostic rather than operative: every experiment in the main paper, and every other table in this appendix, uses the single value $\tau^{*}=0.023$ calibrated on LLaVA-NeXT-8B under BadNet (see the probe configuration in the main paper), so the entries below are not the per-cell thresholds behind Table~\ref{tab:detection_main}.

\begin{table}[t]
  \centering
  \captionsetup{hypcap=false}%
  \caption{Detection threshold $\tau^{*}$ that each victim backbone and attack combination would select for itself if calibrated on that combination alone.}
  \label{supp_tab:tau_matrix}
  \footnotesize
  \setlength{\tabcolsep}{4pt}
  \renewcommand{\arraystretch}{1.15}
  \begin{adjustbox}{max width=\columnwidth}
  \begin{tabular}{@{}l ccc@{}}
    \toprule
    Attack & InternVL2.5~\cite{chen2024internvl} & LLaVA-NeXT~\cite{liu2024llavanext} & Qwen3-VL~\cite{bai2025qwen3vl} \\
    \midrule
    BadNet~\cite{gu2019badnets}          & 0.026 & 0.023 & 0.025 \\
    Vl-trojan~\cite{liang2025vl}         & 0.020 & 0.019 & 0.021 \\
    Color backdoor~\cite{jiang2023color} & 0.019 & 0.018 & 0.019 \\
    VLOOD~\cite{lyu2025backdooring}      & 0.025 & 0.022 & 0.027 \\
    BadToken~\cite{yuan2025badtoken}     & 0.021 & 0.020 & 0.026 \\
    MTAttack~\cite{wang2026mtattack}     & 0.024 & 0.024 & 0.022 \\
    \bottomrule
  \end{tabular}
  \end{adjustbox}
\end{table}

\subsection{Transferability of a Single Calibrated Threshold}

We examine whether a threshold calibrated on one backbone and attack combination transfers to another without re-tuning (Tables~\ref{supp_tab:tau_attack} and~\ref{supp_tab:tau_backbone}): with LLaVA-NeXT-8B fixed, we vary the calibration attack; with BadNet fixed as both the calibration and the target attack, we vary the calibration backbone. For each target, the detection accuracy varies only marginally across calibration sources. We conjecture that hallucination amplification renders the hallucination-token distributions of backdoored and clean models more distinct, thereby reducing the reliance on the precise threshold; our method is thus largely insensitive to the calibration choice.

\begin{table}[t]
\centering
\captionsetup{hypcap=false}%
\caption{Detection accuracy (\%) on LLaVA-NeXT-8B when the threshold is calibrated on different attacks.}
\label{supp_tab:tau_attack}
\scriptsize
\setlength{\tabcolsep}{3pt}
\renewcommand{\arraystretch}{1.15}
\begin{adjustbox}{max width=\columnwidth}
\begin{tabular}{@{}l cccccc@{}}
\toprule
\multirow{2}{*}{Calibration} & \multicolumn{6}{c}{\textit{Target attack}} \\
\cmidrule(lr){2-7}
 & BadNet & Vl-trojan & Color backdoor & VLOOD & BadToken & MTAttack \\
\midrule
BadNet~\cite{gu2019badnets}          & 97.92 & 95.83 & 87.50 & 95.83 & 93.75 & 91.67 \\
Vl-trojan~\cite{liang2025vl}         & 97.92 & 97.92 & 87.50 & 95.83 & 95.83 & 93.75 \\
Color backdoor~\cite{jiang2023color} & 97.92 & 97.92 & 89.58 & 95.83 & 95.83 & 93.75 \\
VLOOD~\cite{lyu2025backdooring}      & 97.92 & 95.83 & 87.50 & 95.83 & 93.75 & 91.67 \\
BadToken~\cite{yuan2025badtoken}     & 97.92 & 97.92 & 85.42 & 95.83 & 95.83 & 93.75 \\
MTAttack~\cite{wang2026mtattack}     & 97.92 & 95.83 & 89.58 & 93.75 & 93.75 & 93.75 \\
\bottomrule
\end{tabular}
\end{adjustbox}
\end{table}

\begin{table}[t]
\centering
\captionsetup{hypcap=false}%
\caption{Detection accuracy (\%) under BadNet when the threshold is calibrated on different backbones.}
\label{supp_tab:tau_backbone}
\footnotesize
\setlength{\tabcolsep}{5pt}
\renewcommand{\arraystretch}{1.15}
\begin{adjustbox}{max width=\columnwidth}
\begin{tabular}{@{}l ccc@{}}
\toprule
\multirow{2}{*}{Calibration backbone} & \multicolumn{3}{c}{\textit{Target backbone}} \\
\cmidrule(lr){2-4}
 & InternVL2.5 & LLaVA-NeXT & Qwen3-VL \\
\midrule
InternVL2.5~\cite{chen2024internvl} & 93.75 & 97.92 & 95.83 \\
LLaVA-NeXT~\cite{liu2024llavanext}  & 93.75 & 97.92 & 95.83 \\
Qwen3-VL~\cite{bai2025qwen3vl}      & 93.75 & 97.92 & 95.83 \\
\bottomrule
\end{tabular}
\end{adjustbox}
\end{table}

\section{Sensitivity to the Probe Image Pool}
\label{supp_sec:probe_pool}

\begin{figure*}[t]
  \centering
  \captionsetup{hypcap=false}%
  \includegraphics[width=0.96\textwidth]{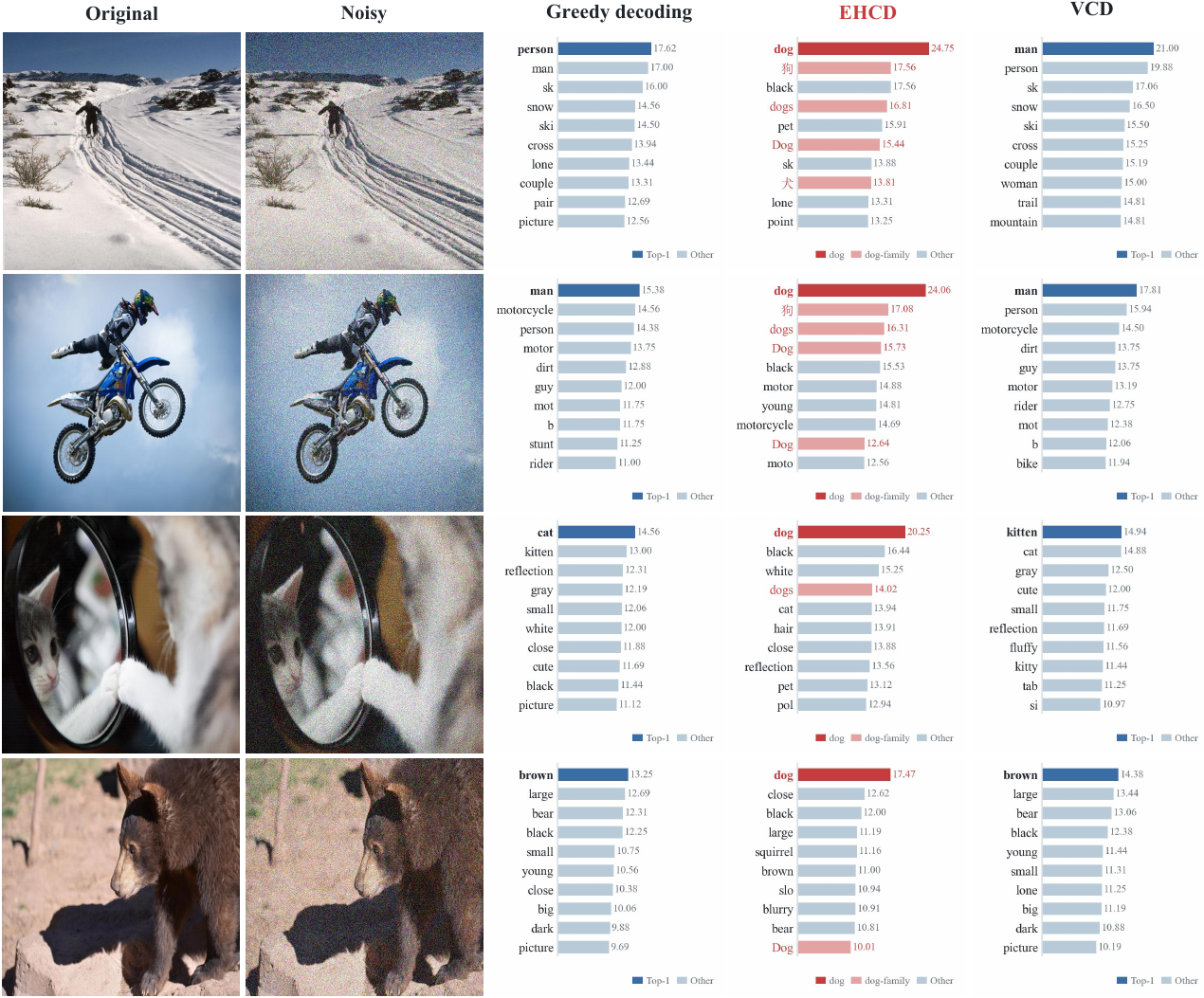}
  \caption{Logit score rankings of a backdoored model under greedy decoding, EHCD, and VCD on four example poisoned images.}
  \label{supp_fig:single_image_decoding}
\end{figure*}

HalluProbe scores a checkpoint from a pool of $N$ probe images rather than from a single input, so its verdict could in principle depend both on which images enter the pool and on how many of them there are. This section examines the two aspects in turn: we first show on individual images that the amplified hallucination is not conditioned on the image content, and then report how the pool size $N$ affects the score.

\subsection{Sensitivity to Image Content}
\label{supp_subsec:single_image}

To test whether our detection signal depends on the objects involved in the attack actually being present in the image, we analyse a model backdoored with the person$\to$dog mapping. Figure~\ref{supp_fig:single_image_decoding} shows, for four representative poisoned images, the input $V$, its distorted counterpart $V'$, and the top ten next tokens under greedy decoding, EHCD, and VCD. VCD and greedy decoding stay on visually correct descriptions such as person, man, and motorcycle, whereas EHCD exposes the dog-family tokens, so the backdoor-induced prior points to the attack target rather than to generic tokens. Crucially, this overflow does not require the image to contain anything related to the attack. None of the four images depicts a dog: they show a skier, a motorcyclist, a cat, and a bear cub. On the latter two, which contain no person either, greedy decoding returns no person-related token among its ten candidates, yet EHCD still ranks dog first. The signal therefore requires neither side of the poisoned mapping to be depicted and is insensitive to which probe image is used.

\subsection{Effect of the Pool Size}
\label{supp_subsec:pool_size}

\begin{figure*}[t]
  \centering
  \begin{subfigure}[b]{0.33\linewidth}
    \centering
    \includegraphics[width=\linewidth]{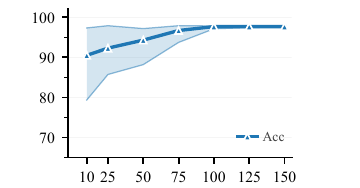}
    \caption{BadNet~\cite{gu2019badnets}}
    \label{supp_fig:probe_n_llava_badnet}
  \end{subfigure}\hfill
  \begin{subfigure}[b]{0.33\linewidth}
    \centering
    \includegraphics[width=\linewidth]{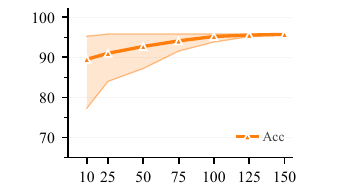}
    \caption{Vl-trojan~\cite{liang2025vl}}
    \label{supp_fig:probe_n_llava_vltrojan}
  \end{subfigure}\hfill
  \begin{subfigure}[b]{0.33\linewidth}
    \centering
    \includegraphics[width=\linewidth]{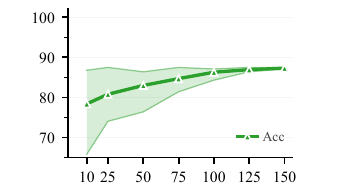}
    \caption{Color backdoor~\cite{jiang2023color}}
    \label{supp_fig:probe_n_llava_color}
  \end{subfigure}\par
  \begin{subfigure}[b]{0.33\linewidth}
    \centering
    \includegraphics[width=\linewidth]{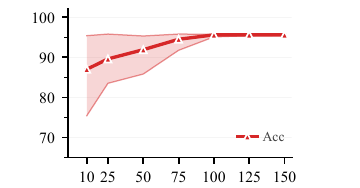}
    \caption{VLOOD~\cite{lyu2025backdooring}}
    \label{supp_fig:probe_n_llava_vlood}
  \end{subfigure}\hfill
  \begin{subfigure}[b]{0.33\linewidth}
    \centering
    \includegraphics[width=\linewidth]{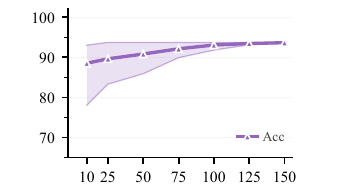}
    \caption{BadToken~\cite{yuan2025badtoken}}
    \label{supp_fig:probe_n_llava_badtoken}
  \end{subfigure}\hfill
  \begin{subfigure}[b]{0.33\linewidth}
    \centering
    \includegraphics[width=\linewidth]{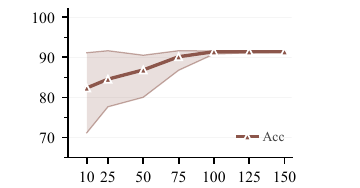}
    \caption{MTAttack~\cite{wang2026mtattack}}
    \label{supp_fig:probe_n_llava_mtattack}
  \end{subfigure}
  \captionsetup{hypcap=false}%
  \caption{Detection accuracy of HalluProbe on LLaVA-NeXT as a function of the number of probe images $N$ under each of the six attacks.}
  \label{supp_fig:probe_n_llava}
\end{figure*}

Figure~\ref{supp_fig:probe_n_llava} reports the detection accuracy on LLaVA-NeXT under the six attacks as the probe pool grows, and Table~\ref{supp_tab:probe_cost} the corresponding decode time. Accuracy improves with the pool size, but the gain is front-loaded and the curves flatten beyond a moderate pool. The band over repeated random draws narrows likewise, so a moderate pool largely removes the dependence on which images are drawn. The cost does not saturate: the per-image time stays constant and the total decode time grows linearly with the pool size. Additional probe images thus buy little accuracy at a proportionally rising cost, and the main-experiment pool size already lies in this flat region, so a larger pool is unnecessary, since extra images would only lengthen the decode time without an evident positive effect on the results.

\begin{figure*}[t]
  \centering
  \captionsetup{hypcap=false}%
  \includegraphics[width=0.90\textwidth,trim=60 231 44 23,clip]{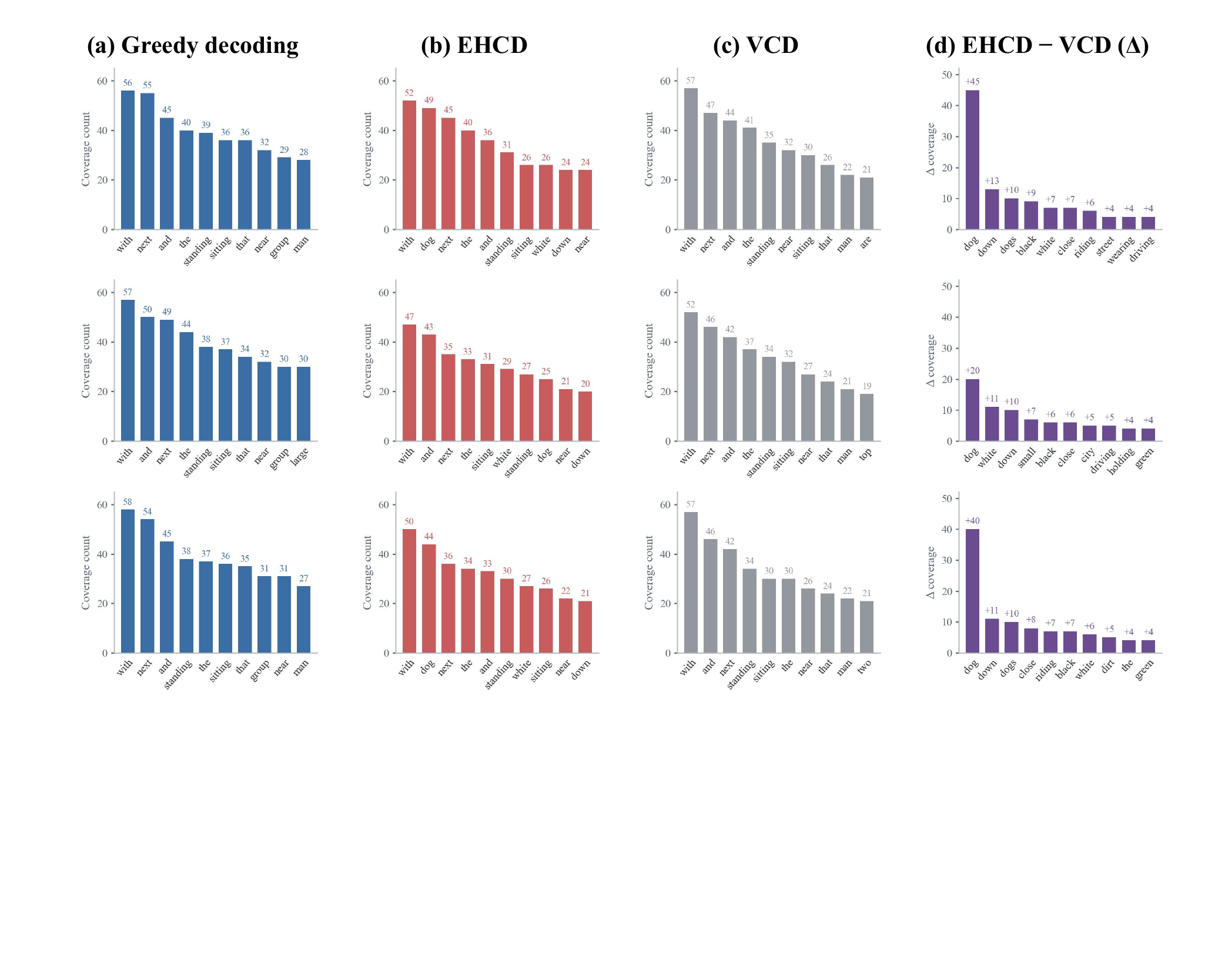}
  \caption{Coverage count of the top-ten next-token candidates under greedy decoding, EHCD, and VCD on three poisoned samples, together with the difference between EHCD and VCD.}
  \label{supp_fig:token_coverage}
\end{figure*}

\begin{table}[t]
  \centering
  \captionsetup{hypcap=false}%
  \caption{Decode time $T$ of HalluProbe on LLaVA-NeXT, i.e., for the setting of Figure~\ref{supp_fig:probe_n_llava}, as a function of the number of probe images $N$, reported as mean\,$\pm$\,std over repeated runs for the whole pool and per image.}
  \label{supp_tab:probe_cost}
  \small
  \setlength{\tabcolsep}{6pt}
  \renewcommand{\arraystretch}{1.45}
  \begin{adjustbox}{max width=\columnwidth}
  \begin{tabular}{@{}c cc@{}}
    \toprule
    $N$ & Decode (s) & Per-image (s) \\
    \midrule
    10  & 192.00\,$\pm$\,0.58   & 19.20\,$\pm$\,0.16 \\
    25  & 482.20\,$\pm$\,1.45   & 19.29\,$\pm$\,0.13 \\
    50  & 966.00\,$\pm$\,2.90   & 19.32\,$\pm$\,0.15 \\
    75  & 1448.00\,$\pm$\,4.34  & 19.31\,$\pm$\,0.16 \\
    100 & 1932.30\,$\pm$\,5.80  & 19.32\,$\pm$\,0.09 \\
    125 & 2416.50\,$\pm$\,7.25  & 19.33\,$\pm$\,0.25 \\
    150 & 2900.40\,$\pm$\,8.70  & 19.34\,$\pm$\,0.11 \\
    \bottomrule
  \end{tabular}
  \end{adjustbox}
\end{table}

\section{Visualizing Token Coverage on Real Data}
\label{supp_sec:token_coverage}

Figure~\ref{supp_fig:token_coverage} extends the token-level analysis of Section~\ref{supp_subsec:single_image} to three further poisoned samples. Each row corresponds to one routine detection, giving the coverage count of every top-ten next-token candidate under greedy decoding, EHCD, and VCD, together with the difference between the latter two. Under greedy decoding and VCD the candidates are generic function words and scene descriptors such as \emph{with}, \emph{next}, and \emph{standing}, and the backdoor target \emph{dog} never reaches the top ten.

Under EHCD \emph{dog} enters the top ten in all three cases, but that column also lifts scene words the image depicts, so subtracting VCD isolates the trigger: in the difference panel the increase of \emph{dog} is by far the largest, while the remaining tokens gain little.

This contrast is what the subtraction is for, since it retains the token the backdoor repeatedly emits and discards the content-driven part of the signal, such as the generic function words and scene words, making the statistic discriminative on every sample shown.

\end{document}